\documentclass[prb, aps, twocolumn, floatfix, superscriptaddress]{revtex4-1}
\usepackage{amssymb}
\usepackage{amsmath}
\usepackage{graphicx}
\usepackage[english]{babel}
\usepackage[latin1]{inputenc}

\begin{document}

\title{Cavity Optomechanics with Polariton Bose-Einstein Condensates}

\author{D. L. Chafatinos}
\affiliation{Centro At\'omico Bariloche \& Instituto Balseiro (C.N.E.A.) and CONICET, 8400 S.C. de Bariloche, R.N., Argentina.}

\author{A.~S. Kuznetsov}
\affiliation{Paul-Drude-Institut f\"{u}r Festk\"{o}rperelektronik, Leibniz-Institut im Forschungsverbund Berlin e.V., Hausvogteiplatz 5-7,\\ 10117 Berlin, Germany.}

\author{S. Anguiano}
\affiliation{Centro At\'omico Bariloche \& Instituto Balseiro (C.N.E.A.) and CONICET, 8400 S.C. de Bariloche, R.N., Argentina.}

\author{A.~E. Bruchhausen}
\affiliation{Centro At\'omico Bariloche \& Instituto Balseiro (C.N.E.A.) and CONICET, 8400 S.C. de Bariloche, R.N., Argentina.}

\author{A.~A. Reynoso}
\affiliation{Centro At\'omico Bariloche \& Instituto Balseiro (C.N.E.A.) and CONICET, 8400 S.C. de Bariloche, R.N., Argentina.}

\author{K. Biermann}
\affiliation{Paul-Drude-Institut f\"{u}r Festk\"{o}rperelektronik, Leibniz-Institut im Forschungsverbund Berlin e.V., Hausvogteiplatz 5-7,\\ 10117 Berlin, Germany.}

\author{P.~V. Santos}
\affiliation{Paul-Drude-Institut f\"{u}r Festk\"{o}rperelektronik, Leibniz-Institut im Forschungsverbund Berlin e.V., Hausvogteiplatz 5-7,\\ 10117 Berlin, Germany.}

\author{A. Fainstein}
\email[Corresponding author, e-mail: ]{afains@cab.cnea.gov.ar}
\affiliation{Centro At\'omico Bariloche \& Instituto Balseiro (C.N.E.A.) and CONICET, 8400 S.C. de Bariloche, R.N., Argentina.}

\date{\small December 23rd, 2019}

\begin{abstract}
We report the experimental study of a hybrid quantum solid state system comprising two-level artificial atoms coupled to cavity confined optical and vibrational modes. In this system combining cavity quantum electrodynamics and cavity optomechanics, excitons in quantum wells play the role of the two-level atoms and are strongly coupled to the optical field leading to mixed polariton states. The planar optical microcavities are laterally microstructured, so that polaritons can be confined in wires, 3D traps, and arrays of traps, providing an additional tuning degree of freedom for the polariton energies. Upon increasing the non-resonant laser excitation power, a Bose-Einstein condensation of the polaritons is observed. Optomechanical induced amplification type of experiments with an additional weak laser probe clearly identify the coupling of these Bose-Einstein condensates to 20~GHz breathing-like vibrations confined in the same cavities. With single continuous wave non-resonant laser excitation, and once the laser power overpasses the threshold for Bose-Einstein condensation in trap arrays, mechanical self-oscillation similar to phonon ``lasing'' is induced with the concomitant observation of Mollow-triplet type mechanical sidebands on the Bose-Einstein condensate emission. High-resolution spectroscopic photoluminescence experiments evidence that these vibrational side-band resolved lines are enhanced when neighboring traps are red-detuned with respect to the BEC emission at overtones of the fundamental 20~GHz breathing mode frequency. These results constitute the first demonstration of coherent cavity polariton optomechanics and pave the way towards a novel type of hybrid devices for quantum technologies
, phonon lasers, and phonon-photon bidirectional translators.
\end{abstract}


\maketitle

Hybrid devices composed of different physical components with complementary functionalities constitute one trending line of research for new quantum and communication network technologies, for signal storage, processing, conversion, and transmission.\cite{Kurizki} Cavity optomechanics\cite{ReviewCOM} constitutes one domain in which hybrid designs have been envisaged both for the test of fundamental quantum physics at the mesoscopic level as well as for novel devices.\cite{Rogers} In cavity optomechanics photons are confined and strongly coupled to vibrational degrees of freedom thus leading, under the appropriate conditions, to dynamical back-action phenomena including optically induced coherent self-oscillation of the mechanical mode\cite{Kippenberg,Grudinin} and, conversely,  laser cooling of the mechanical mode, even down to the quantum ground state.\cite{O'Connell,Teufel,Chan,Verhagen} One application of cavity optomechanical devices is in the bidirectional conversion between signals of contrastingly different frequency range, for example between classical microwave and optical light,\cite{Bochmann,Bagci} even with prospects for the transfer of quantum states.\cite{Andrews} Cavity polaritons, the strongly coupled quantum states combining an exciton (as a two-level artificial atom) and a cavity confined photon, are the fundamental excitations of other hybrid cavity quantum electrodynamic system comprising semiconductor optical microcavities with embedded quantum wells.\cite{Weisbuch} Polaritons display both the delocalized nature of the photonic wavefunction and the finite mass of the excitons, plus non-linearities arising from the Coulomb interaction via the excitonic fraction of the state.\cite{Bajoni} Since the initial report of cavity polaritons in semiconductor microcavities, their Bose-Einstein condensation,\cite{Kasprzak} superfluidity,\cite{Amo} lasing also under electrical pumping,\cite{Schneider} and multistable behavior\cite{Rodriguez} have been demonstrated.

\begin{figure}[!hht]
    \begin{center}
    \includegraphics[trim = 0mm -20mm 0mm 0mm, clip=true, keepaspectratio=true, width=\columnwidth, angle=0]{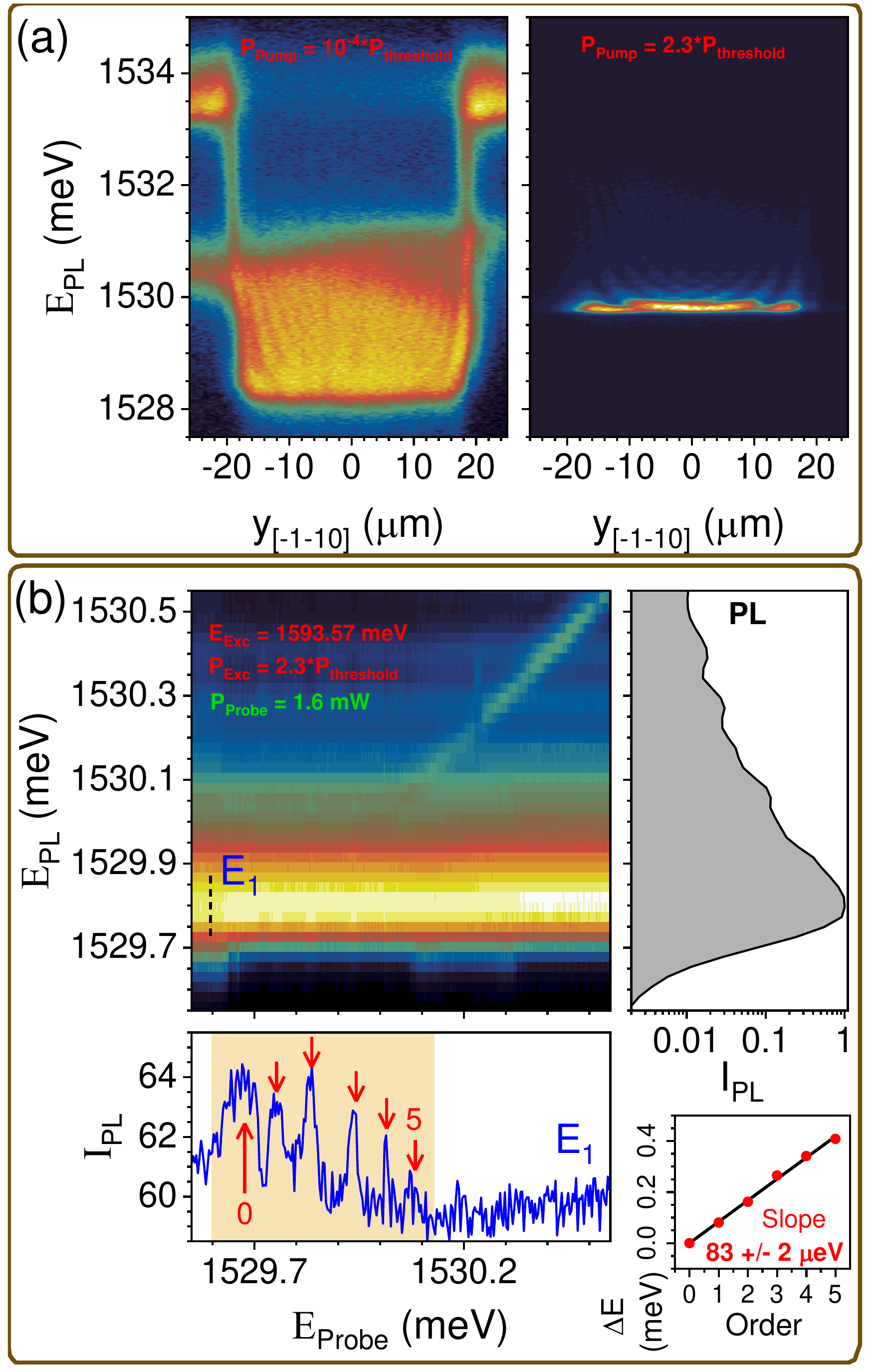}
    \end{center}
    \vspace{-1.6 cm}
\caption{
\textbf{Bose-Einstein condensate optomechanical induced amplification.}
Panel (a) shows photoluminescence color maps of exciton-polaritons confined in a  $\sim 30~\mu$m stripe. Two spatial images are shown, for non-resonant excitation powers well below ($P_{in}=10^{-4} P_{Th}$) and above($P_{in}=2.3 P_{Th}$) the threshold power $P_{Th}$.
Panel (b) presents a photoluminescence (PL) color map as a function of the probe laser energy, for the polariton stripe. The right panel displays the integrated PL intensity. The bottom left panel shows the amplitude of the fundamental state BEC emission as a function of the probe laser energy. The peaks corresponding to OMIA processes are highlighted, and their energies as a function of the peak order are shown in the bottom right panel.
\label{fig1}}
\end{figure}

Hybrid quantum systems combining both cavity quantum electrodynamics and cavity optomechanics have been theoretically proposed,\cite{Restrepo,Kyriienko} with predictions of cooling at the single-polariton level, peculiar quantum statistics, and coupling to mechanical modes of both dispersive and dissipative nature leading to unconventional bistable behavior. Cavity optomechanics with a polariton Bose-Einstein condensate opens intriguing perspectives, particularly in view of the potential access to an optomechanical strong-coupling regime, and the possibility to use vibrations to actuate on such a macroscopic quantum fluid. Indeed, the strength of the optomechanical interaction in conventional photon cavity optomechanics is measured by the optomechanical cooperativity, $C=\frac{4g_0^2n_{cav}}{\kappa \Gamma_m}$, where $g_0$ is the single-photon optomechanical coupling factor, $n_{cav}$ is the number of photons in the cavity, and $\kappa$  and $\Gamma_m$ are the photon and mechanical decay rates, respectively.\cite{ReviewCOM} When a Bose-Einstein condensate of polaritons is considered, one can expect strongly enhanced values of the cooperativity $C$ due to the very large coherent population of the condensate, coherence times that can be two orders of magnitude larger than the photon lifetime in the cavity, and optomechanical coupling interactions mediated by the excitons that can be resonantly enhanced.\cite{Jusserand}
Such hybrid optomechanical systems based on  Bose-Einstein condensate have, to the best of our knowledge, so far only been demonstrated with cold atoms in cavities.\cite{Brennecke} It is the purpose of this paper to investigate the rich physics emerging from the coupling of a polariton Bose-Einstein condensate in a semiconductor microcavity, with the ultra-high frequency vibrations confined in these resonators.\cite{FainsteinPRL2013,Anguiano}

The studied device consists of polaritons in arrays of $\mu$m-sized intracavity traps created by patterning the spacer of an (Al,Ga)As microcavity in-between steps of the molecular beam epitaxy growth process.\cite{Kuznetsov}  Mesas with a nominal height of 12 nm of different shapes in the exposed spacer layer induce a lateral modulation of the cavity thickness and, therefore, of the photonic cavity energy in the final structure. The etching depth results in a blue-shift of the optical cavity mode in the etched areas by 9~meV (4.5~nm) with respect to the non-etched regions.  This lateral modulation was used to create confinement in 2D (wires and stripes), 3D (dots), as well as dot arrays consisting of non-etched areas surrounded by etched barriers. These polariton traps were studied by low-temperature reflection and photoluminescence (PL) in Ref.\,[\onlinecite{Kuznetsov}]. The sample is in the strong coupling regime both in the etched and non-etched regions, leading to microcavity polaritons in these two regions with different energies and photon/exciton contents. Bose-Einstein condensates (BECs) can be efficiently induced in the traps both with non-resonant and resonant excitation. As reported previously for similar planar\cite{FainsteinPRL2013} and pillar\cite{Anguiano} microcavities, these structures also confine breathing-like vibrations of longitudinal character (polarized along the growth direction $z$) with fundamental frequency around $\nu_m^0 \sim 20$~GHz, and even order overtones of this fundamental mode with frequencies $\nu_m^n = (1+2n) \nu_m^0$.  We will concentrate here in experiments performed on a stripe of $30~\mu$m thickness (very weak lateral confinement), and an array of small coupled square traps of $1.6~\mu$m size separated by $2~\mu$m etched regions (see a detailed description of the structure in the supplementary material).

\begin{figure*}[!hht]
    \begin{center}
    \includegraphics[trim = 0mm -20mm 0mm 0mm, clip=true, keepaspectratio=true, width=1.8\columnwidth, angle=0]{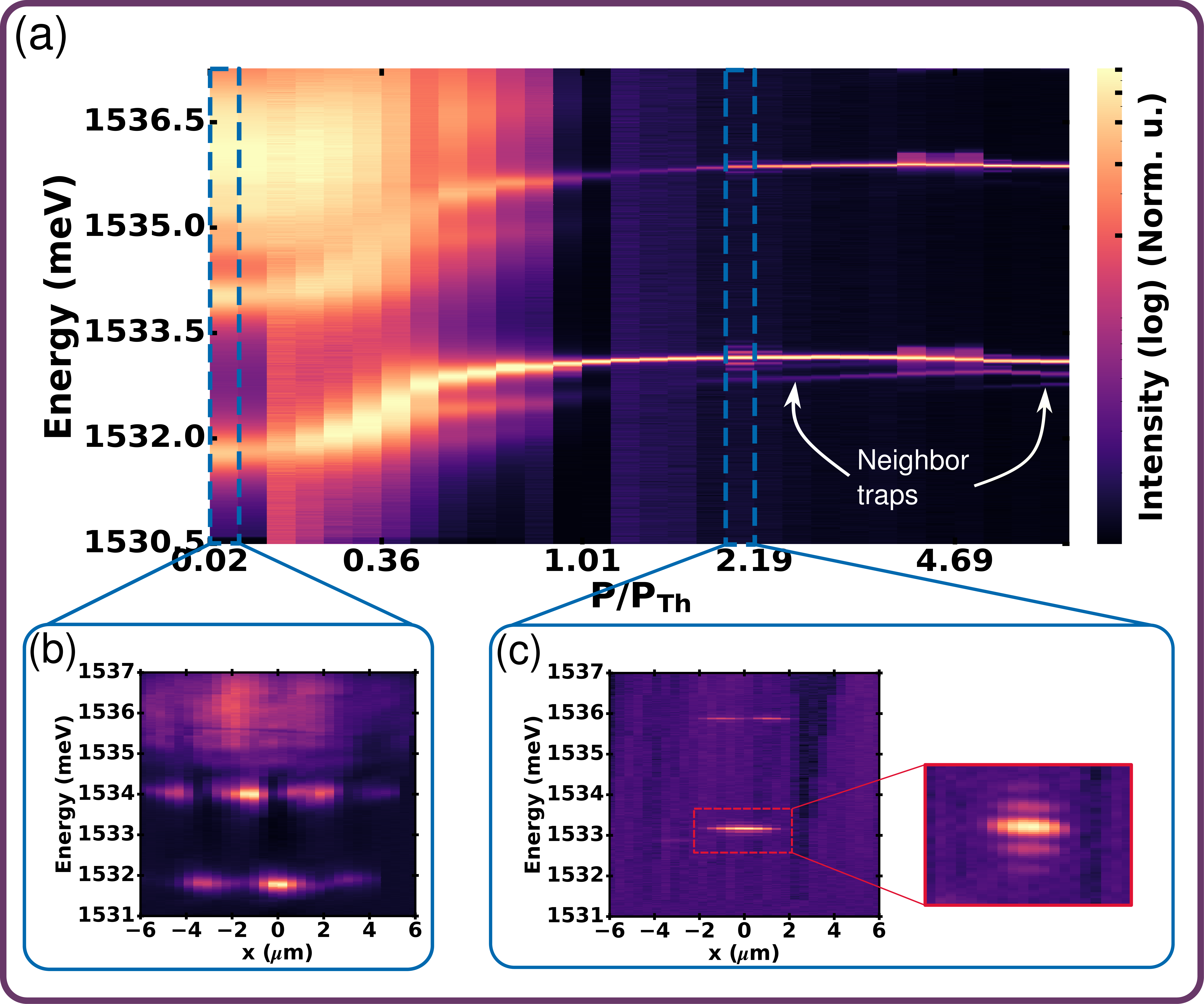}
    \end{center}
    \vspace{-0.8 cm}
\caption{
{\bf Bose-Einstein condensation in a polariton trap array.}
Panel (a) presents a photoluminescence color map of exciton-polaritons confined in a 3D trap array of squares of nominal size $1.6~\mu$m.
The non-resonant excitation power dependence of the photoluminescence is shown. The excitation power is given in terms of the threshold power $P_{Th} \sim 19$mW. For clarity each spectrum has been normalized to its maximum amplitude, and a logarithmic amplitude scale is used. Both emission from the pumped trap and the red-shifted neighbor traps can be clearly observed. Note the appearance of low and high energy sidebands on the BEC state for certain pump powers. The figure also includes photoluminescence spatial images for two situations, below (pabel (b)) and above (panel (c)) the Bose-Einstein condensation threshold. For the latter a zoom of the central pumped trap is included, evidencing the appearance of the mentioned well resolved sidebands. In panel (b) the neighbor traps can be clearly identified at $x \sim \pm 3.6 \mu$m.
\label{fig2}}
\end{figure*}

Panel (a) of Fig.~\ref{fig1} shows the  spectrally and spatially resolved PL image of the $30~\mu$m-wide polariton stripe recorded at 5~K. The cavity mode in this microstructure is slightly negatively detuned with respect to the QW heavy-hole exciton state. The spectra were obtained with non-resonant excitation using a Ti-Sapphire \emph{cw} laser (760\,nm) with pump powers $P_{Pump}$ below, and above the condensation threshold. At low powers ($P_{Pump}=10^{-4} P_{Th}$, with $P_{Th}$ the threshold power), the different closely packed levels confined due to the lateral confinement in the stripe can be clearly identified. Above threshold  ($P_{Pump}= 2.3 P_{Th}$), Bose-Einstein condensation of the polaritons is evidenced by the narrowing and concentration of the emission at the bottom of the trap, as well as by the blue-shift of the states induced by the Coulomb interaction associated to the excitonic-fraction of the polaritons.

Figure~\ref{fig1}(b) presents results of a two-laser optomechanically induced amplification (OMIA) experiment,\cite{Renninger} performed on this $30~\mu$m-thick stripe. The first (pump) laser creates a BEC in the stripe. The second weak probe laser was incident with an angle of 20$^\circ$, and was scanned from below to above the condensate energy (a detail of the experimental set-up is provided in the supplementary material).  The color map in Fig.~\ref{fig1}(b) shows the PL intensity as a function of the probe laser energy. The right panel presents the integrated intensity, evidencing the structure of the closely packed states of the polariton stripe.
OMIA is evidenced by the variation of the luminescence intensity at the peak of the BEC emission, as a function of the probe laser energy as also displayed at the bottom of Fig.~\ref{fig1}(b). Notably, a series of peaks separated by the energy of the fundamental breathing-like cavity confined vibrational mode ($\sim 20$\,GHz$\sim 83\,\mu$eV) appears at the high energy side of the BEC peak. OMIA with the final (control) state corresponding to the polariton fundamental energy is conceptually similar to a stimulated Raman process. The spectra in Fig.~\ref{fig1}(b) shows that first order and several higher order Stokes replicas (creation of one or several phonons) are amplified, but anti-Stokes ones (annihilation of a phonon) are not. The origin of this asymmetry resides in the double optical resonance condition satisfied for the Stokes processes, but not for the anti-Stokes ones.\cite{FainsteinPRL1995} In fact, due to the finite incidence angle, the probe beam can enter the cavity and resonantly couple to modes with energy higher than that of the BEC. On the contrary, no modes are available at energies below the BEC, thus strongly suppressing this channel. The observation of several replicas of the 20~GHz phonon assisted transfer of energy from the probe laser to the BEC evidences a strikingly efficient optomechanical interaction.

To exploit such efficient polariton-vibrational coupling for optomechanical back-action with the BEC, we propose a double resonant scheme involving confined states of  sites of an  array of $1.6~\mu$m-wide square traps separated by $2\mu$m thick barriers. Conceptually, the idea follows the phonon-laser scheme described in Ref.~[\onlinecite{Grudinin}]. Here, Grudinin \textit{et al.} used two separate optical resonators that are spatially approached so that the coupling-induced splitting becomes resonant with a vibrational state of the system. Under this resonant condition, pumping with a $cw$ laser at the high-energy optical mode of the coupled cavities, induces a non-linear build-up of the phonon population and a transition to a self-oscillation regime (equivalent to a parametric phonon laser).  The threshold condition to attain zero effective mechanical damping and regenerative self-oscillation of the phonon mode is given by the optomechanical cooperativity being $C>1$.\cite{ReviewCOM,Renninger} We next discuss  how this can be accomplished in a BEC array, coupled to the confined vibrations evidenced in the OMIA experiments discussed above.

Figure~\ref{fig2} presents through a color-map the pump-power dependence of the PL intensity of the $1.6~\mu$m trap array. Spatially resolved PL images below and above threshold power are also presented. The experiments were performed at 5~K with non-resonant excitation (760~nm) using a $cw$ Ti-Sapphire Spectra Physics Matisse laser, with a microluminescence set-up based on a $\times20$ microscope objective (NA=0.3, spot size $\sim 3~\mu$m), a high-resolution triple additive spectrometer (resolution $\sim 0.15$cm$^{-1} \sim 20~\mu$eV), and a liquid-N$_2$ cooled CCD (see the supplementary material for further details).  The laser spot addresses mainly a single trap, but neighboring traps can be excited through the tails of the laser Gaussian spot, and also due to lateral propagation of the excitons in the reservoir and polaritons in the traps.  The fundamental and first excited states of the pumped trap, as well as weaker contributions from neighbor traps, can be identified in the color map and spatial images in Fig.~\ref{fig2}. With increasing pump power the polariton modes blue-shift, and above a threshold value the intensity of the fundamental mode increases nonlinearly attaining the Bose-Einstein condensation. Concomitantly with the nonlinear amplitude increase, the emission linewidth narrows strongly. The measured linewidth is limited by the resolution of the triple additive spectrometer. Measurements using a custom-made tandem Fabry-Perot-triple additive spectrometer,\cite{RozasRSI} allow to access the true-linewidth, showing that the longest coherence time, observed at approximately 37~mW of pump power, is $\sim 530$~ps (linewidth $\sim 8~\mu$eV). That is, it is two orders of magnitude larger than the polariton lifetime measured at low powers ($\sim 6-10$~ps). Note that the neighbor traps also blue-shift with increasing power, though with a weaker slope, thus attaining a power dependent red-shift with respect to the pumped trap. This feature will become of critical relevance in what is discussed next.

\begin{figure*}[!hht]
    \begin{center}
    \includegraphics[trim = 0mm -30mm 0mm 0mm, clip=true, keepaspectratio=true, width=1.9\columnwidth, angle=0]{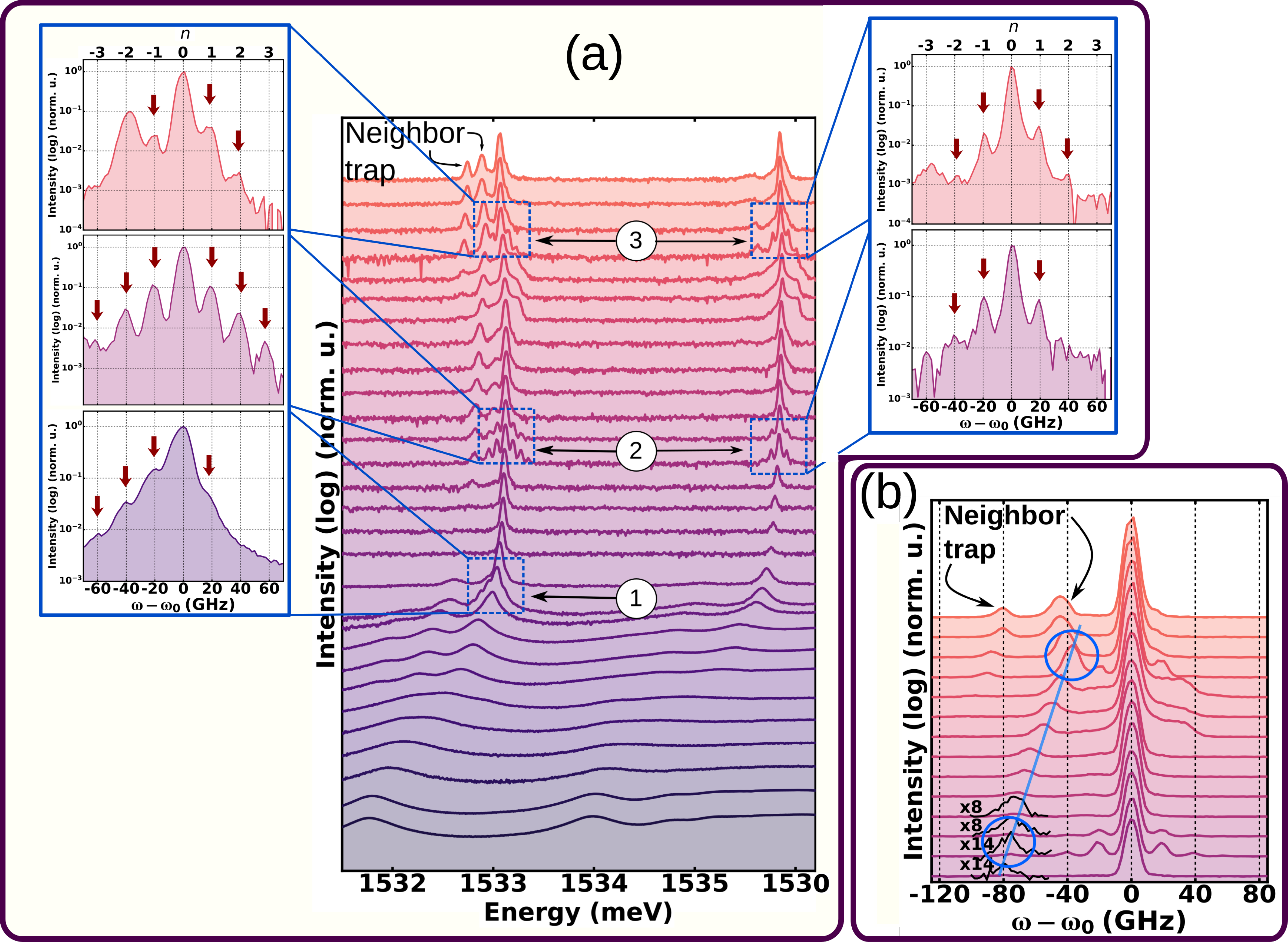}
    \end{center}
    \vspace{-0.8 cm}
\caption{
{\bf Regenerative mechanical self-oscillation induced by a BEC.}
Panel (a): Photoluminescence spectra for varying non-resonant excitation powers, for a $1.6~\mu$m polariton square trap array. The numbered arrows highlight the BEC optomechanical signatures: arrow 1) equally spaced low energy sidelines; arrow 2) clear well resolved sidebands at both sides of the fundamental BEC and first excited polariton peaks; and arrow 3) reappearance of the sidebands both for the fundamental and first excited polariton trap states. A detail of some of the spectra in regions (1-3) corresponding to the fundamental BEC and first excited polariton state, highlighting the energy scale of the sidebands coincident with the confined mechanical vibration $\nu_m^0=20$GHz$\sim 83~\mu$eV, is included.
Panel (b): Detail of higher power photoluminescence spectra, shifted in energy and shown relative to the fundamental BEC state. The red-shifted neighbor trap photoluminescence contributions are highlighted. Circular labels indicate the situation when the neighbor traps detuning is 2 and 4 times the confined mechanical vibration $\nu_m^0=20$GHz.
\label{fig3}}
\end{figure*}

The spatial images in Fig.~\ref{fig2} also show that, precisely when the linewidth of the BEC emission is smallest, the fundamental mode develops well resolved equally separated side-bands, both on the high and low energy sides of the main peak. Such kind of sidebands are a signature of a coherent modulation of the BEC emission, as previously observed for example for narrow emitters (semiconductor quantum-dots and diamond NV-centers) externally driven by surface acoustic waves (SAWs).\cite{QDs,NVs} The condition for the observation of these well-resolved sidebands is that the lifetime of the emitter is longer than the period of the modulation (or equivalently, that the modulation frequency is larger than the linewidth of the modulated line). The intensity of the sidebands, in contrast, reflects the magnitude of the driving modulation.  What is particularly noteworthy in the results displayed by the spatial image in  Fig.~\ref{fig2}, is that in our experiments there is no external harmonic driving, only a $cw$ laser is used to non-resonantly excite the semiconductor QWs.

Figure~\ref{fig3}(a) shows the high-resolution spectra (in log intensity scale) for the full scan of measured powers, corresponding to the polariton trap-array PL color map shown in Fig.~\ref{fig2}. For clarity, and because of the large variation in absolute intensity, each spectrum is normalized to its BEC maximum peak amplitude.  Again the blue-shift and narrowing of the main BEC peak on increasing power can be clearly observed, together with the appearance of red-shifted neighbor traps, particularly once the condensation in the central pumped trap is attained. The arrows with labels 1-3 in Fig.~\ref{fig3}(a) identify spectra we want to highlight: arrow 1) intense and equally spaced low energy sidelines appear, reminiscent of phonon assisted PL; arrow 2) on increasing power, clear well-resolved sidebands appear on both sides of the main BEC peak, and this is also observed for the emission corresponding to the first excited trap polariton level; and arrow 3) on increasing power, and after having disappeared, the sidebands reappear both for the fundamental and first excited polariton trap states.

\begin{figure}[!hht]
    \begin{center}
    \includegraphics[trim = 0mm -10mm 0mm 0mm clip=true, keepaspectratio=true, width=0.9\columnwidth, angle=0]{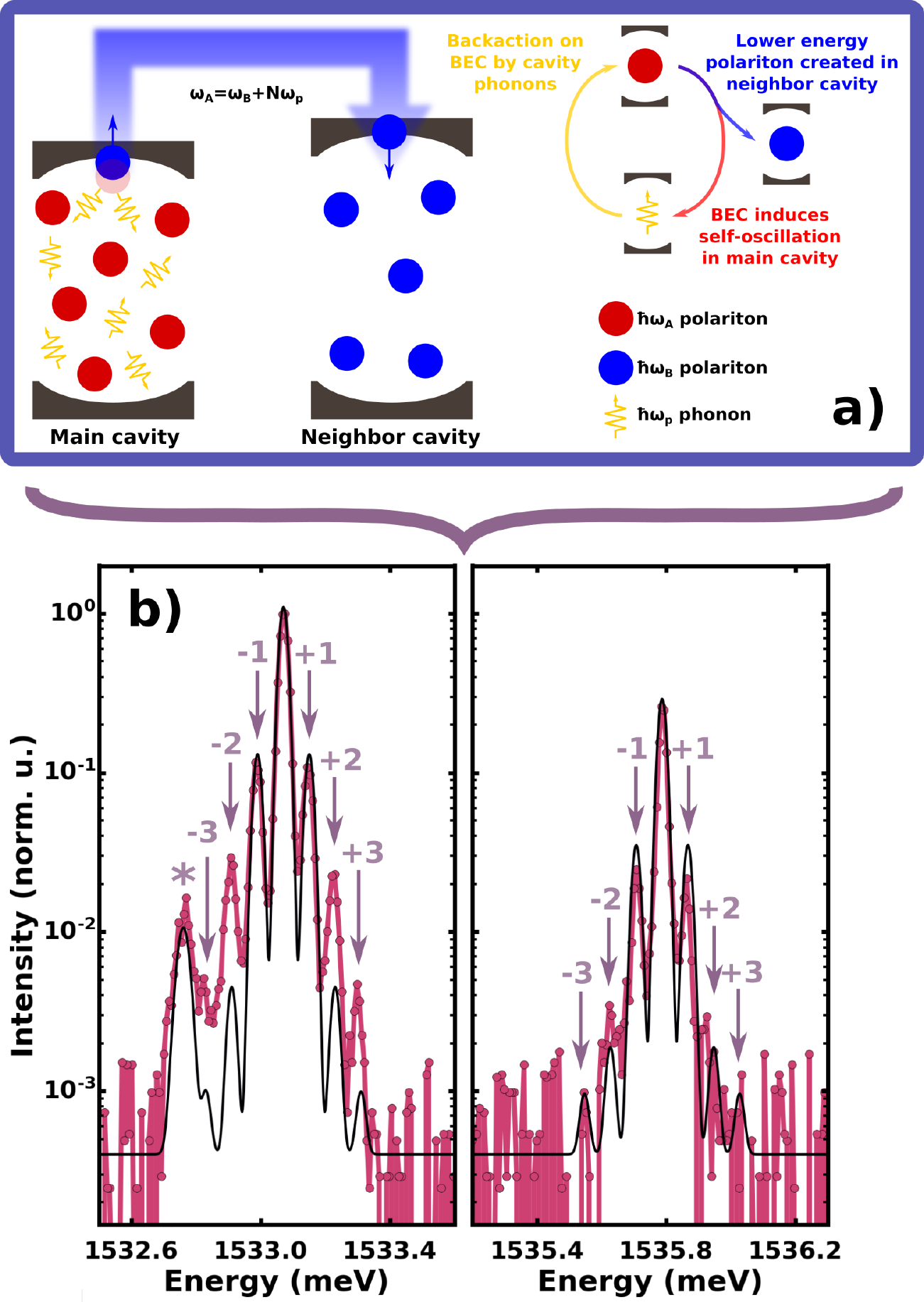}
    \end{center}
    \vspace{-0.8 cm}
\caption{
{\bf BEC optomechanics: self-oscillation and mechanical induced BEC sidebands.}
Panel (a) shows a scheme of the proposed mechanism. i) The BEC couples to the neighbor traps inducing regenerative optomechanical self oscillation at the frequency of the confined mechanical breathing fundamental vibration $\nu_m^0=20$GHz. And ii) the generated coherent oscillations back-act on the BEC and the excited polariton states, inducing well resolved sidebands. Panel (b) is a comparison of measured BEC spectra in the presence of mechanically induced sidebands, with the model of Eq.\,(\ref{FrequencyModulation}) (see text for details). For the calculation we used $\chi\sim 0.65$. The asterisk indicates a peak due to photoluminescence from a neighbor trap, which was added ad-hoc in the model.
\label{fig4}}
\end{figure}

Panel (a) of Fig.\ref{fig3} presents a detail of these relevant spectra, highlighting that the sidebands correspond precisely to equally spaced secondary peaks separated by the energy of the fundamental cavity confined breathing mode ($\nu_0 \sim 20$~GHz$\sim 83~\mu$eV). In the presence of a coherent harmonic driving of the BEC, the spectrum of the luminescence is expected to be proportional to $P[\omega]$:\cite{QDs}
\begin{equation}
P[\omega]=\sum_{n=-\infty}^{\infty} \frac{J_n^2(\chi)}{[\omega-(\omega_0-n\,\omega_d)]^2+\gamma^2},
\label{FrequencyModulation}
\end{equation}
that is, a sum of Lorentzians with full-width-at-half-maximum (FWHM) of $2\gamma$, weighted by squares of Bessel functions $J^2_n(\chi)$. Here the argument $\chi$  is a dimensionless parameter expressing the frequency shift induced by the harmonic driving on the BEC resonance ($\Delta \omega_0$), in units of the driving frequency $\omega_d$ ($\chi=\Delta \omega_0/\omega_d$). The Lorentzians have maxima at frequencies $\omega = \omega_0-n \omega_d$, where $n$ is an integer. An example of $P[\omega]$ for $\chi=0.65$ is shown in Fig.~\ref{fig4}, highlighting the similarity with the measured spectral shapes.  Comparison with the measured magnitude of the sidebands allows for an estimation of the energy shift induced by the cavity mechanical coherent vibrations. From the magnitude of the main peak and that of the first sidebands, we estimate $\Delta E_{BEC}\sim 55~\mu$eV. Using this value, through a calculation of the polariton energy dependence on strain (deformation potential) and cavity thickness (interface displacement), we estimate the average number of phonons ($\langle N\rangle$) and strain associated to the regenerative self-oscillation induced by the BEC. We obtain $\langle N\rangle \sim 2 \times 10^5$ (note that the thermal occupation of this mode at 5~K is $\langle N\rangle_{\text{Thermal}} \sim 5$), corresponding to a maximum breathing of the cavity of $\sim 660$\,pm, implying a strain of around $0.1\%$ (see the supplementary material for details on these calculations).

Self-oscillation in cavity optomechanics is attained when the external \textit{cw} laser excitation is blue-shifted with respect to the cavity mode (in the one-mode situation), or when the higher cavity mode is excited (in the two-mode situation).\cite{Renninger} The polariton BEC as an internal coherent source cannot thus, by itself, induce self-oscillation, because both Stokes and anti-Stokes processes would be equally probable. This symmetry is broken through the coupling of the BEC to the red-shifted neighbor traps. Figure~\ref{fig3}(b) shows details of the spectra recorded with the highest powers in panel (a), presented as a function of the detuning respect to the BEC emission. It becomes evident that, indeed, clear resolved sidebands appear precisely when one of the neighbor traps is red-detuned by integer numbers of $\nu_m^0 \sim 20$~GHz. The spectra highlighted with arrow 1 in Fig.~\ref{fig2}(a), not shown in panel (b), also appears when the neighbor trap is detuned close to $\delta \nu = -6 \nu_m^0$.  The experimental observations can then be understood as follows (see the scheme in Fig.~\ref{fig4}): i) the optomechanical interaction pumped by the BEC and coupled to a precisely red-detuned neighbor trap, induces self-oscillations at the fundamental frequency of the 20~GHz mode, and ii) these coherent oscillation  back-acts by modulating the BEC, thus leading to the observed sidebands.
A closer look at Fig.~\ref{fig3}(b) shows, in addition, that the intensity of the phonon side-band increases when the neighbor trap is red-detuned by even numbers of  $\nu_m^0 \sim 20$~GHz: $\delta \nu = -2 \nu_m^0=-40$~GHz, $\delta \nu = -4 \nu_m^0=-80$~GHz, added to $\delta \nu = -6 \nu_m^0$ as mentioned above. A red-detuning of $\delta \nu=-3\nu_m^0$, which is coincident with the first order of $-\nu_m^1=-60$~GHz induces, in contrast, very strong spectral changes of the main BEC peak. The latter seem to reflect a multistability behavior and, thus, interaction strengths beyond the weak perturbation described by Eq.\,\eqref{FrequencyModulation}.~\cite{Note} A full description of this regime is presently not available.

Polaritons exert force by radiation pressure (through their photonic fraction) and by electrostriction (deformation potential photoelastic interaction mediated by the excitonic component). That is, the optomechanical coupling factor is  $g_0=S_x\,g_0^{ph} + S_c\,g_0^{RP}$, with $S_x (S_c)$ the exciton(photon) fraction of the polariton mode, and $g_0^{ph}(g_0^{RP})$ the photoelastic(radiation pressure) contribution to the optomechanical coupling factor (see the discussion of this point in the supplementary material). The excitons in the QWs could lead to a resonantly enhanced optical force.\cite{Jusserand}
In the studied sample, however, the QWs are positioned at the anti-nodes of the optical field in the unstructured regions to optimize the polariton strong coupling. The phonon strain vanishes in these positions, so that the deformation potential coupling is not expected to contribute significantly. A small departure of this strain cancellation might be present in the trap arrays due to the lateral confinement of the phonon field. Besides, evaluation of these forces shows that radiation pressure couples mainly to the fundamental 20~GHz mode, while electrostriction does with the overtone mode at $\sim 60$~GHz (see the supplementary material). Because in the described experiments the sidebands reflect an harmonic modulation with a frequency of 20~GHz, we will consider in the following only the coupling through radiation pressure. We estimate next the conditions for BEC-induced self-oscillation in the studied devices.

The number of polaritons in the BEC can be estimated as $n_{pol}=\eta \frac{P_{Pump}}{\hbar \omega} \tau$. Here $\eta$ is the fraction of excited electron-hole pairs that condense from the exciton reservoir into the BEC. $P_{Pump}$ is the non-resonant laser power, $\hbar \omega$ the laser photon energy, so that $\frac{P_{Pump}}{\hbar \omega}$ is the number of excitons generated by the non-resonant excitations.  And $\tau$ is the exciton reservoir lifetime. From time-resolved differential reflectivity measurements, we estimate $\tau\sim 2$\,ns. With this value we get $n_{pol}\sim \eta \times 10^7\, P_{Pump}$, with $P_{Pump}$ given in mW. Based on previous studies of cavity confined modes in semiconductor microcavities, we assume the mechanical $Q$-factor is $Q_M \sim 2 \times 10^3$.\cite{Anguiano,Perrin} and thus $\Gamma_M=2\pi \nu_0/Q_M \sim 2\pi \times 10$~MHz. We take as $\kappa$ the BEC decoherence rate, which from the measured coherence time of $\sim 530$~ps ($Q_{BEC} \sim 2 \times 10^5$) is $\kappa \sim 2\pi \times 1.8$~GHz.
The radiation pressure optomechanical coupling factor in a pillar microcavity of similar size as the studied traps has been calculated in Ref.~[\onlinecite{Villafane1}]. In the BEC the photon fraction is very close to 1/2, and thus we take half of the reported value, $g_0^{RP} \sim 2\pi \times 25$~kHz. From the threshold condition $C=1$ one thus obtains that the self-oscillation threshold condition for the BEC is $P_{Th} = \frac{1}{\eta \times 10^7} \frac{\kappa \Gamma_M}{4 (g_0^{RP})^2} \sim \frac{0.7}{\eta}$[mW]. Consequently, assuming that $20 \%$ of the photoexcited electron-hole pairs end up populating the BEC, this estimation gives $P_{Th} \sim 4$~mW. This implies that the hybrid polariton BEC optomechanical system is already in conditions of self oscillation at the powers required for condensation (Bose-Einstein condensation is observed above $\sim 19$~mW for the $1.6~\mu$m traps under the experimental conditions used).  Consequently, self-oscillation should be observed whenever the conditions for double resonance are satisfied, i.e. that neighbor traps are red-detuned from the BEC integer numbers of $\nu_m^0$, as is indeed experimentally observed.

We have demonstrated coherent phonon generation and self-modulation at ultra-high vibrational frequencies using a polariton Bose-Einstein condensate. Different from other proposals that have demonstrated laser-induced regenerative self-oscillation of a phonon mode,\cite{ReviewCOM,Kippenberg,Grudinin} in our scheme this is driven by an internally emitting polariton source, which could, in principle, be electrically pumped.\cite{Schneider} The coherent phonons generated are efficiently emitted into the supporting substrate,\cite{Huyhn} thus providing a new platform for an electrically driven parametric phonon laser. The observed spectrally resolved vibrational sidebands, on the other hand, are a manifestation of coupled tripartite atom-cavity-mechanics polarons, as proposed by Restrepo and  coworkers.\cite{Restrepo} The cavity quantum electrodynamics feature of the system results in the strong energy conserving coupling between photons and excitons, while the optomechanics radiation pressure term couples off-resonant mechanical and photonic modes of widely different frequencies (GHz and hundreds of THz, respectively). Thus the demonstrated technology could be at the base of novel technologies for frequency conversion between light and mechanical (or microwave) signals in the 20~GHz range.\cite{Bochmann,Bagci,Andrews,Rueda} Similarly, our findings demonstrate that mechanical vibrations can be used to coherently actuate on a macroscopic quantum fluid (the BEC) efficiently and at very high frequencies (frequencies larger than the decoherence time of the quantum state). The strength of the coupling could be enhanced by several orders of magnitude, as compared to the demonstrated radiation pressure interactions, exploiting the resonant character of the photoelastic coupling mediated by excitons in QWs.\cite{Jusserand} This could be attained by engineering the position of the QWs so as to optimize simultaneously the light-exciton (cavity quantum electrodynamic) and the exciton-phonon (cavity optomechanical) interactions. By doing this, in addition, coupling to cavity mechanical modes of hundreds of GHz could be performed,\cite{Villafane2} thus providing access to operation and signal transduction at the so-called extremely high frequency range.

\section{Acknowledgments}

We thank Evelyn~G.~Coronel for helping with the experiments in the initial stages of this project.
We acknowledge partial financial support from the ANPCyT-FONCyT (Argentina) under grants PICT1015-1591 and PICT2015-1063, from the  German Research Foundation (DFG) under grant 359162958, and the joint Bilateral Cooperation Program between the German Research Foundation (DFG) and the Argentinian Ministry of Science and Technology (MINCyT) and CONICET.





\onecolumngrid
\newpage
\setcounter{figure}{0}  
\renewcommand{\figurename}{FIG. S\!\!}
\renewcommand{\thefigure}{\arabic{figure}}

\newcommand{\NN}{\mathbb{N}}
\newcommand{\RR}{\mathbb{R}}
\newcommand{\ZZ}{\mathbb{Z}}
\newcommand{\bea}{\begin{eqnarray}}
\newcommand{\eea}{\end{eqnarray}}
\newcommand{\bei}{\begin{itemize}}
\newcommand{\eei}{\end{itemize}}
\newcommand{\be}{\begin{equation}}
\newcommand{\ee}{\end{equation}}
\newcommand{\bse}{\begin{subequations}}
\newcommand{\ese}{\end{subequations}}
\newcommand{\bfg}{\begin{figure}}
\newcommand{\efg}{\end{figure}}
\newcommand{\e}{\epsilon}

\newcommand{\vzero}{\boldsymbol{0}}
\newcommand{\ve}{\boldsymbol{e}}
\newcommand{\vk}{\boldsymbol{k}}
\newcommand{\vp}{\boldsymbol{p}}
\newcommand{\vq}{\boldsymbol{q}}
\newcommand{\vrr}{\boldsymbol{r}}
\newcommand{\vx}{\boldsymbol{x}}
\newcommand{\vy}{\boldsymbol{y}}
\newcommand{\vF}{\boldsymbol{F}}
\newcommand{\veta}{\boldsymbol{\eta}}
\newcommand{\veps}{\boldsymbol{\epsilon}}
\newcommand{\vtheta}{\boldsymbol{\theta}}
\newcommand{\vomega}{\boldsymbol{\omega}}
\newcommand{\vphi}{\boldsymbol{\varphi}}
\newcommand{\vPsi}{\boldsymbol{\Psi}}
\newcommand{\Hop}{\hat{H}}
\newcommand{\Sop}{\mathbf{\hat{S}}}
\newcommand{\Svec}{\mathbf{S}}

\newcommand{\photonOp}{\hat{c}}
\newcommand{\phononOp}{\hat{b}_m}
\newcommand{\phononOpD}{\hat{b}_m^\dagger}
\newcommand{\excitonOp}{\hat{e}}
\newcommand{\polaritonOp}{\hat{p}}
\newcommand{\excitonNOp}{e}
\newcommand{\polaritonNOp}{p}
\newcommand{\phononNOp}{b_m}
\newcommand{\phononNOpEq}{b_{m,\mathrm{eq}}}

\newcommand{\phononFreq}[1]{\omega_m^{#1}}
\newcommand{\ci}{\mathrm{i}}

\newcommand{\traptop}{1}
\newcommand{\traplow}{2}


\begin{center}
\underline{\huge{\textbf{Additional Information}}}
\end{center}
\vspace{0.5cm}

These supplementary material includes I) a description of the studied polariton microstructures, II) the experimental set-ups, III) some details of the laser power dependence of the amplitude and linewidth of the
polartion emission in the $1.6 \mu$m square trap array, IV) a brief description of the method used to calculate the radiation pressure contribution to $g_0$, V) a description of the method used to evaluate the displacement and average number of phonons associated to the coherent self-oscillation induced by the coupling of the BEC state with the neighbor traps, VI) the calculation of the vibrational spectra expected for both radiation pressure and electrostriction optical forces, and last, VII) a brief description of the Hamiltonian describing the two-polariton system coupled to the cavity confined vibrations.

\section{Studied structure}

\begin{figure}[!hht]
    \begin{center}
    \includegraphics[trim = 0mm -20mm 0mm 0mm,scale=0.50,angle=0]{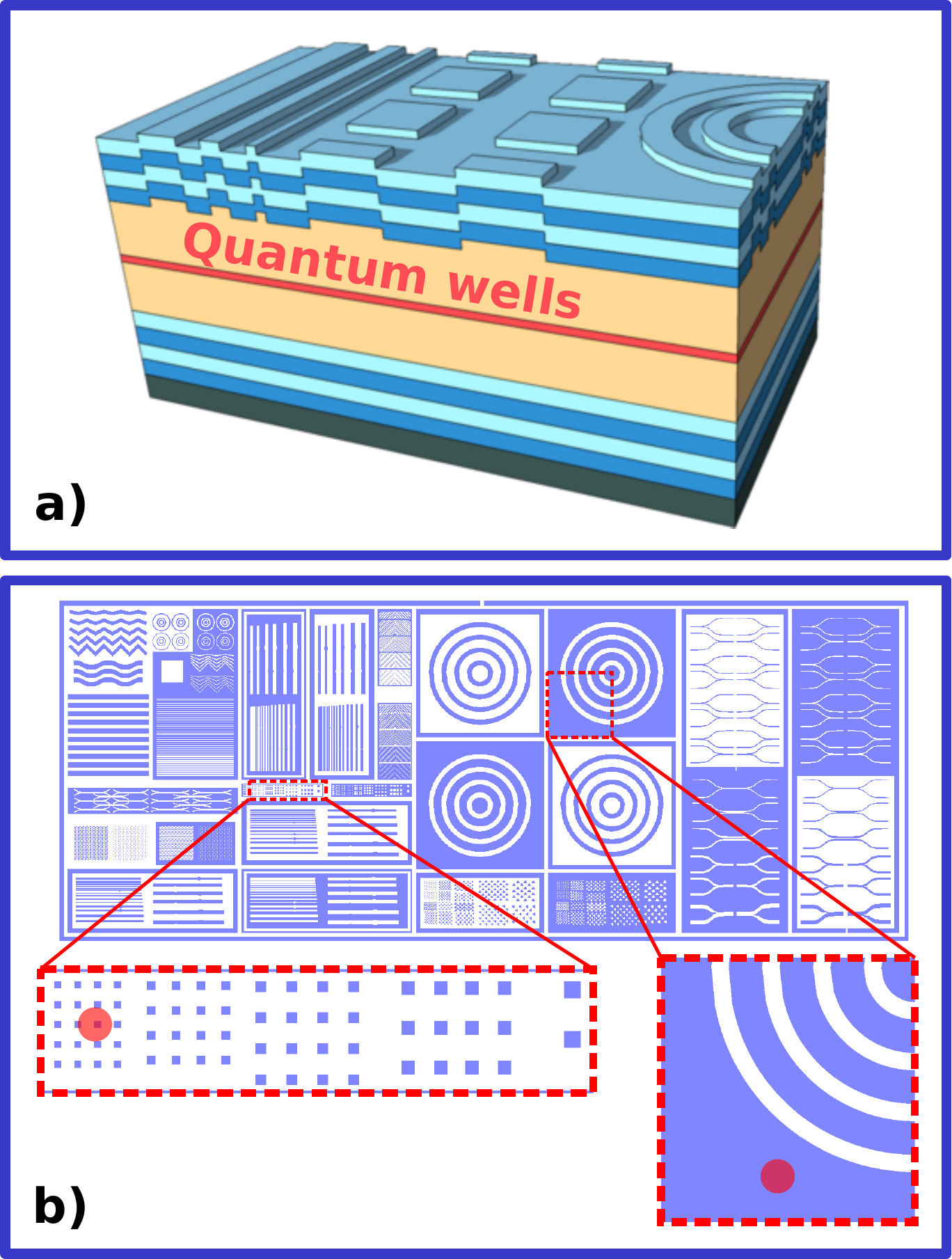}
    \end{center}
    \vspace{-0.8 cm}
\caption{
{\bf Studied microstructured planar microcavity.}
Panel (a) presents a sketch of the (Al,Ga)As microcavity with the structured spacer enclosing GaAs QWs grown on a GaAs(001)
substrate. The thickness of the spacer between the distributed Bragg reflectors (DBRs) was varied by combining etching and overgrowth by
molecular beam epitaxy, resulting in regions with different polariton energies. Panel (b) shows a detail of the two microstructures
studied in this work, an extended region similar to a stripe of $\sim 30 \mu$m thickness, and an array of $1.6 \mu$m square traps separated by $2 \mu$m edged barrier regiones. The red circles mark approximately the regions probed in the reported experiments.
\label{Structure}}
\end{figure}

The studied device consists of polaritons in arrays of $\mu$m-sized intracavity traps created by patterning an (Al,Ga)As microcavity in-between growth steps by molecular beam epitaxy (see a scheme in Fig.~S\ref{Structure}).\cite{Kuznetsov}  First a 4.43-$\mu$m thick lower distributed Bragg reflector (DBR) consisting of 36 $\lambda$/4 ($\lambda$ is the optical wavelength) pairs of Al$_{0.15}$Ga$_{0.85}$As/Al$_x$Ga$_{1-x}$As with the Al composition $x$ continuously reducing from 0.80 in the first stack to 0.45 in the last stack. The first 120 nm of the Al$_{0.30}$Ga$_{0.70}$As microcavity spacer were then deposited including six 15-nm-thick GaAs quantum wells (QWs) placed at the antinode positions of the
microcavity optical mode. The structure was subsequently capped by a 170-nm-wide Al$_{0.15}$Ga$_{0.85}$As layer spacer. The sample was then taken out of the molecular beam epitaxy (MBE) chamber and then patterned by means of photolithography and wet chemical etching. The latter creates mesas with a nominal height of 12 nm of different shapes in
the exposed spacer layer (see Fig.~S\ref{Structure}), thus inducing a lateral modulation of the cavity thickness and, therefore, of the cavity energy in the final structure. The etching depth results in a blueshift of the optical cavity mode in the etched areas by 9 meV (4.5 nm) with respect to the nonetched regions. The upper surface of the
etched layer corresponds to a node of the optical cavity mode of the whole structure. In this way,  potential impact of roughness or impurities introduced by the ex situ
patterning on optical properties of the structure was minimized. Furthermore, the shallow patterned layer is located more than 140 nm above the QWs, so that they remain
unaffected by the processing. The sample was then reinserted into the MBE system, cleaned by exposure to atomic hydrogen, and overgrown with a
$\lambda$/4 Al$_{0.15}$Ga$_{0.85}$As layer, followed by the upper DBR. The latter consists of 20 $\lambda $/4 pairs of Al$_{0.15}$Ga$_{0.85}$As/Al$_{0.75}$Ga$_{0.25}$As. The sample is in the strong coupling regime both in the etched and nonetched regions, leading to microcavity polaritons in these two regions with different energies and photon/exciton contents.\cite{Kuznetsov} The lateral modulation was used to create 2D (stripes and wires) and 3D (dots) confinement in nonetched areas surrounded by etched barriers, as probed by low-temperature reflection and photoluminescence (PL) in Ref.\cite{Kuznetsov}. Panel (b) in Fig.~S\ref{Structure} illustrates the two microstructured regions in the sample that were studied for the present investigation of polariton BEC optomechanics, a stripe of $30 \mu$m thickness (very weak lateral confinement), and an array of small coupled square traps of $1.6 \mu$m size, separated by $2 \mu$m etched regions.

\section{Experimental set-ups}

Two different set-ups were used, one for the measurement of the two-laser OMIA type of experiments in the $30 \mu$m thickness stripe, the other for the high-resolution optomechanical spectroscopy of the $1.6 \mu$m square trap array.

\begin{figure}[!hht]
    \begin{center}
    \includegraphics[trim = 0mm 0mm 0mm 0mm,scale=0.33,angle=0]{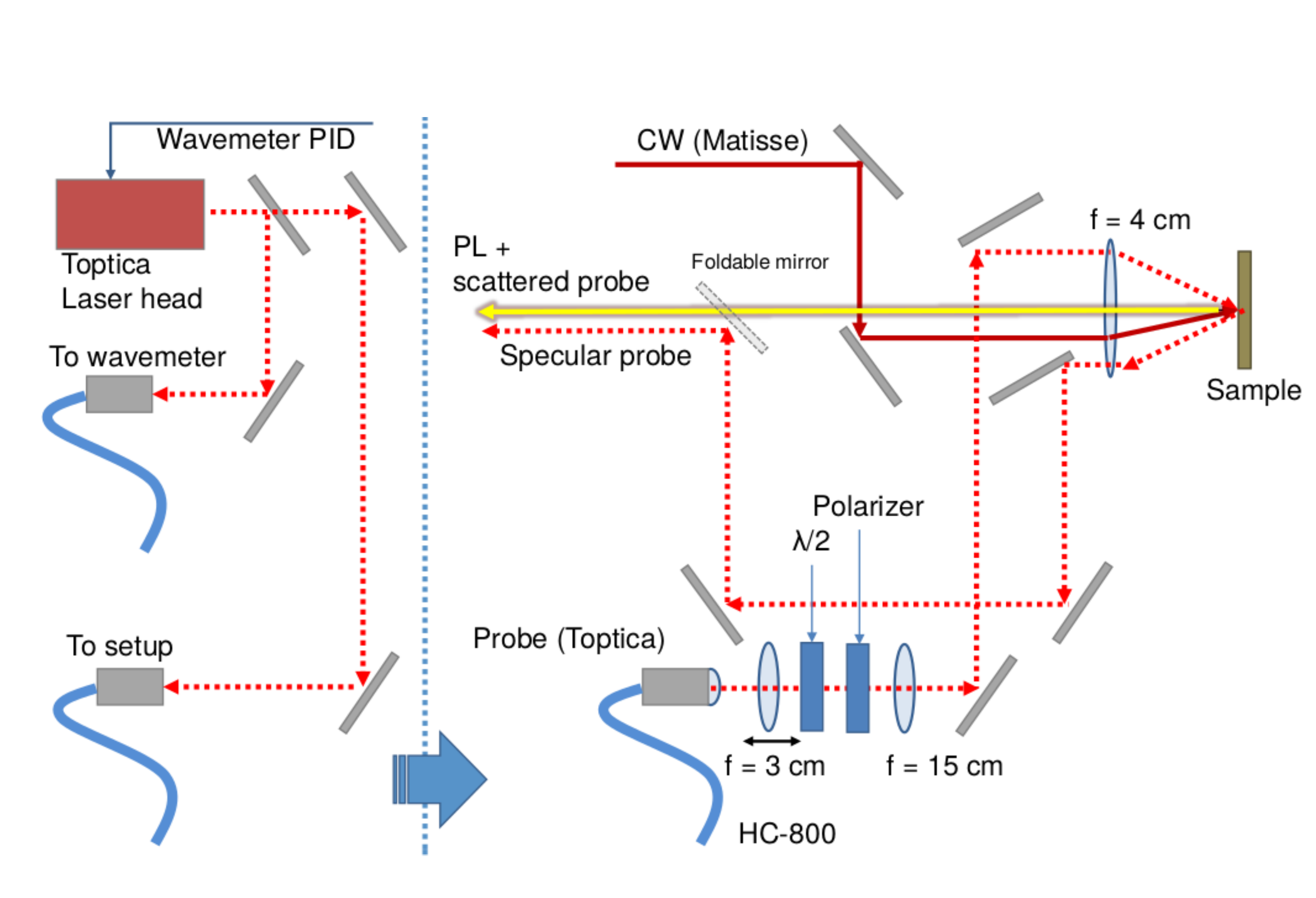}
    \end{center}
    \vspace{-0.8 cm}
\caption{
{\bf Two-laser OMIA setup.}
Scheme of the experimental setup used for the two-laser OMIA type experiments in the $30 \mu$m polariton stripe.
\label{Setup2Laser}}
\end{figure}

Figure~S\ref{Setup2Laser} describes the setup used for the two-laser OMIA type experiments in the $30 \mu$m polariton stripe. A $cw$ Spectra Physics Ti-Sapphire Matisse laser is used for the non-resonant excitation at 760nm. A second weaker Toptica semiconductor stabilized laser, incident with a finite angle, is tuned around the energy of the Bose-Einstein condensate, and light is collected along the normal to the sample.

\begin{figure}[!hht]
    \begin{center}
    \includegraphics[trim = 0mm 0mm 0mm 0mm,scale=0.6,angle=0]{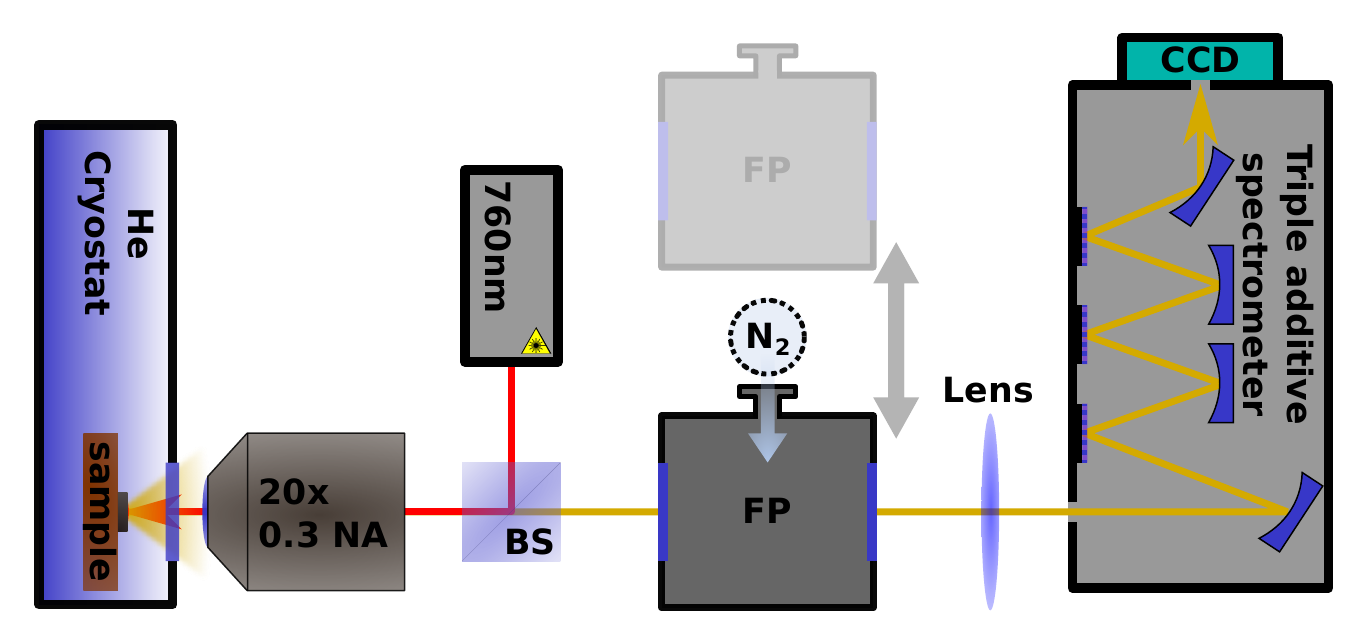}
    \end{center}
    \vspace{-0.8 cm}
\caption{
{\bf High resolution micro-photoluminescence setup.}
Scheme of the experimental setup used for the high-resolution experiments in the array of coupled square traps of $1.6 \mu$m. It comprises a standard microscope photoluminescence configuration with a 20x NA=0.3 objective, a Spectra Physics Ti-Sapphire Matisse laser, and a cold-finger He cryostat, plus two high-resolution features: i) a triple additive T64000 Jobin-Yvon spectometer with 1800 gr/mm gratings, and ii) the possibility to include a home-made tunable Fabry-Perot interferometer in the collection path for sub-pixel resolution (see text for details).
\label{SetupHR}}
\end{figure}

The frequency of the cavity-confined breathing type of mechanical vibrations is large for the typical noise measurements used in lower frequency cavity optomechanics experiments, but rather small for standard optical spectroscopy methods. This imposes strong requirements for the spectral resolution and bandwidth of the vibrational spectroscopy used. To this aim we have used both a photoluminescence microscopy set-up coupled to a triple additive spectrometer (based on three stages of 64~cm each one, and three 1800 gr/mm holographic gratings), and a purposely developed Raman spectroscopy technique based on a tandem Fabry-Perot triple spectrometer multichannel set-up (see the scheme in Fig.~S\ref{SetupHR}).~\cite{RozasRSI} The latter system is composed of a single-pass Fabry-Perot (FP) interferometer coupled to the T64000 Jobin-Yvon triple spectrometer operated in additive configuration.~\cite{RozasRSI}  The light to be analyzed is collected from the sample by a lens, filtered through the FP, and then focused by a second lens into the entrance slit of the spectrometer. The FP contains two high-quality ($\lambda$/200) dielectric mirrors for the near infrared (99$\%$ peak reflectivity centered at 870 nm), which are kept parallel at a fixed distance by three high-quality ($\lambda$/200) cylindrical silica spacers. The mirrors are located in a sealed chamber connected to a pure Nitrogen gas distribution and vacuum system. As the resolution of the spectrometer is better than the free spectral range (FSR) of the FP but not enough to resolve the width of its transmission peaks, the acquired spectrum consists of several broad resolution-limited peaks of which the relevant information is their integrated intensity. By repeating this procedure as a function of the gas pressure, we reconstruct the photoluminescence profile with a sub-pixel resolution improved by almost two orders of magnitude.~\cite{RozasRSI}  The triple spectrometer is equipped with a liquid-N$_2$ cooled charge-coupled device (CCD) multichannel detector which allows for the parallel acquisition of the spectra transmitted through a large set of FP resonances. The excitation is done using a near-infrared Ti:sapphire single-mode Spectra- Physics Matisse TS ring laser, the wavelength of which can be
locked to an external confocal cavity with a precision better than $2 \times
 10^{-6}$ cm$^{-1}$. With this set-up the resolution of the triple spectrometer was improved from  $\sim 0.5$~cm$^{-1} \sim 15$~GHz to $\sim 3 \times 10^{-3}$~cm$^{-1} \sim 90$~MHz.

\section{Power dependence of amplitude and linewidth of the BEC emission}

\begin{figure}[!hht]
    \begin{center}
    \includegraphics[trim = 0mm 0mm 0mm 0mm,scale=0.070,angle=0]{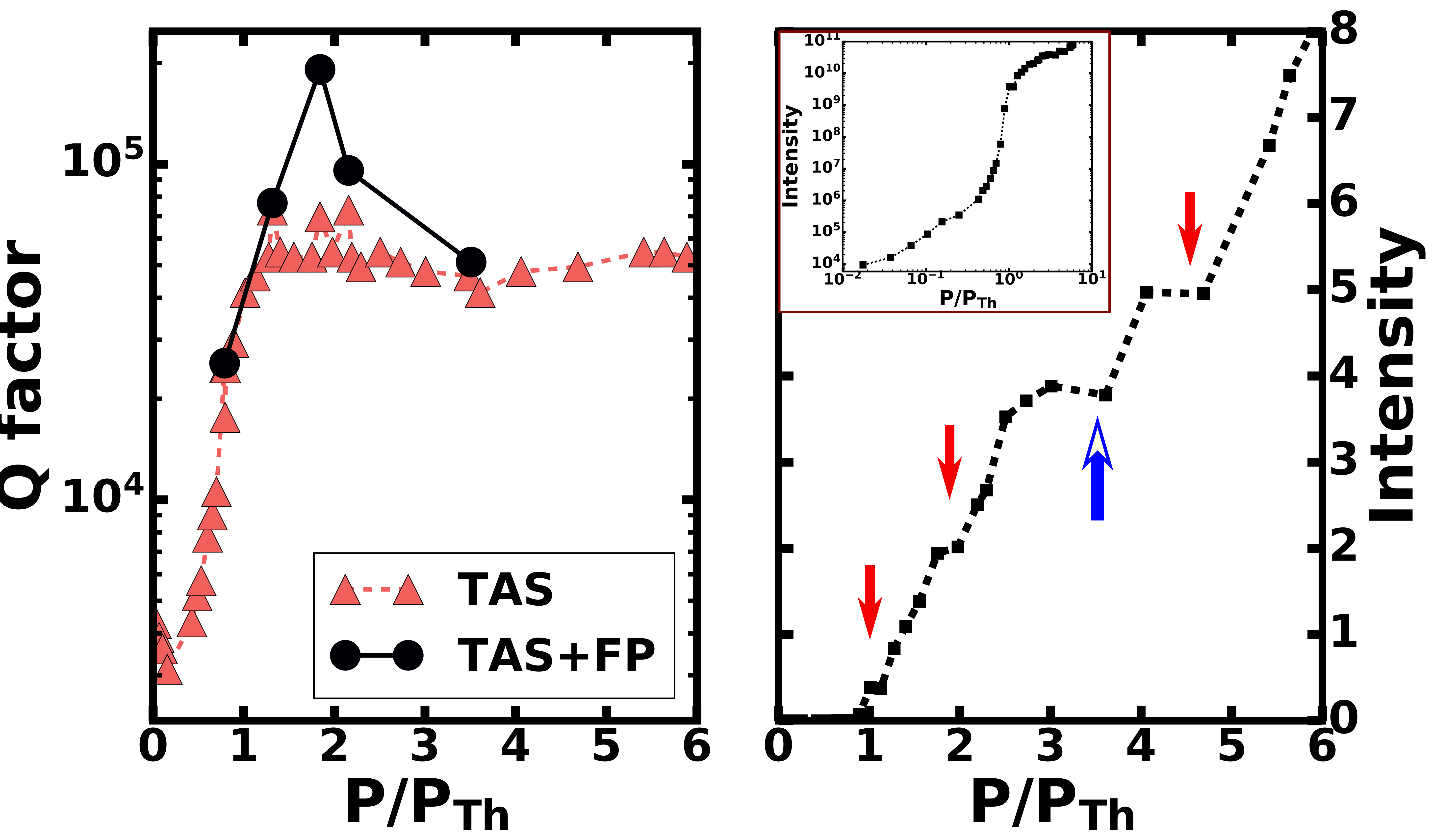}
    \end{center}
    \vspace{-0.8 cm}
\caption{
{\bf Intensity and Q-factor of the BEC emission of the coupled square $1.6 \mu$m array.}
Left panel: Polariton Q-factor determined from luminescence measurements using the high-resolution triple additive spectrometer (TAS, red triangles), and the Fabry-Perot-Triple Additive Spectrometer tandem (TAS+FP, black circles). Above threshold the measurements with the TAS saturate at $Q \sim 5 \times 10^{4}$ due to experimental resolution. The true linewidth (and hence Q-factor) requires the use of the ultra-high-resolution tandem. Right panel: Intensity of the BEC emission as a function of the non-resonant pump power. The inset shows the curve in logarithmic scale to emphasize the non-linear behavior and the threshold power.  Arrows highlight dips that correlate with the regions in which mechanical signatures appear in Fig.3 of the main text.
\label{Intensity}}
\end{figure}

We present in this section some details of the intensity and linewidth of the BEC emission in the array of small coupled square $1.6 \mu$m traps described in Fig.~2 of the main text. The left panel in Fig.~S\ref{Intensity} shows the Q-factor associated to the BEC emission peak, as determined from the luminescence measurements at 5~K using both the high-resolution triple additive spectrometer (TAS, red triangles in the figure), and the ultra-high-resolution Fabry-Perot-Triple-Spectrometer tandem (TAS+FP, black solid circles). The TAS measurements saturate at $Q \sim 5 \times 10^4$ marking the resolution limit of the high-resolution setup ($\sim 20 \mu$eV). The use of the ultra-high-resolution allows to access the true-linewidth ($\sim 8 \mu$eV), corresponding to $Q \sim 2 \times 10^5$ and a BEC coherence time $\tau_{coh} \sim 530$~ps.  The right panel in Fig.~S\ref{Intensity} presents the intensity of the BEC emission peak as a function of the pump power. The inset is displayed in logarithmic scale to emphasize the non-linear dependence and threshold behavior. The main panel present the same results in linear scale, with down red arrows highlighting dips that correlate with the regions in which $\nu_m^0=20$~GHz mechanical sidebands  are evidenced in Fig.3 of the main text. The blue up-arrow corresponds to $-3\nu_m^0=\nu_m^1=-60$~GHz, where also a dramatic change of the BEC emission spectra is apparent in Fig.3 of the main text which, however, has not been yet completely understood. The dips in the BEC emission amplitude evidence a transfer of spectral weight from this mode to the neighbor traps and to the sideband resolved optomechanical secondary peaks.

\section{Calculation of $g_0$}

For the calculation of the radiation pressure contribution to the optomechanical coupling factor $g_{om}^{RP}$, we followed the analysis proposed by Johnson \textit{et al},\cite{Johnson2002} implementing a finite element-method to obtain the electric and acoustic fields. We generalized the approach presented in Refs.\,\onlinecite{Ding2010,Baker} to compute the effects induced by the multiple interfaces at the DBR's boundaries,\cite{Villafane1}
\begin{equation}
\label{geometric_one}
g_{om}^{RP} =  \frac{\omega_c}{2} \sum_i \frac{\oint_{A_i}  (\vec{u} \cdot \hat{n_i}) (\Delta \epsilon_i|\vec{E}_{\|}|^2- \Delta(\epsilon_i^{-1}) |\vec{D}_{\perp}|^2 ) dA_i}{\int \epsilon |\vec{E}|^2 d\vec{r}},
\end{equation}
where $\omega_c$ is the optical angular frequency at resonance, $\vec{u}$ the normalized displacement field, $\hat{n}_i$ the unitary normal-surface vector corresponding to the interface, $\Delta \epsilon_i=\epsilon_{i,left} - \epsilon_{i,right}$ the difference between the dielectric constants of the materials involved, $\Delta \epsilon_i^{-1}=\epsilon_{i,left}^{-1} - \epsilon_{i,right}^{-1}$, $\vec{E}_{\|}$ is the component of the electric-field parallel to the interface surface and $\vec{D}_{\perp}$ is the normal component of the displacement field $\vec{D} = \epsilon_0 \epsilon_r \vec{E}$. The index $i$ runs over every distinct interface $A_i$.

\section{Displacement and average phonon number associated to the regenerative self-oscillation}

\begin{figure*}[!hht]
    \begin{center}
    \includegraphics[trim = 0mm 0mm 0mm 0mm, clip, scale=1.4,angle=0]{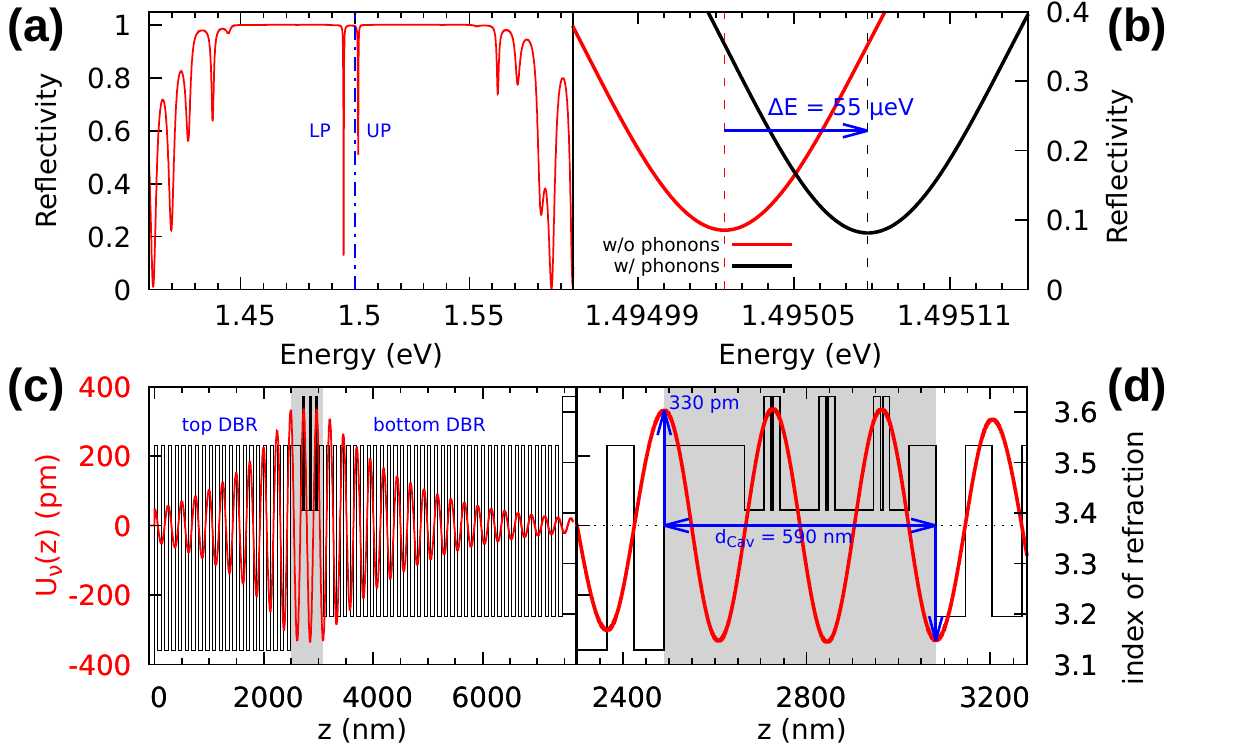}
    \end{center}
    \vspace{-0.5 cm}
\caption{
\textbf{Evaluation of the phonon induced cavity energy shift.}
Panel (a): Calculated optical reflectivity spectrum including an exciton state and a slightly negatively detuned cavity mode. The resulting coupled modes (lower and upper polatitons, LP and UP) are observed.
Panel (b): Estimate LP mode's energy shift matching the experimentally estimated energy shift of $55\mu eV$, corresponding to a coherent self-oscillating vibrational mode with phonon average occupation number $\langle N\rangle\sim 2\times10^5$.
Panel (c): Calculation of the cavity confined fundamental breathing mode at 20~GHz.
Panel (d): Detail  of the cavity spacer's region indicating the vibrational mode that induces a symmetric expansion-contraction of the spacer (vertical blue arrows). The magnitude of the displacement corresponds to $\langle N\rangle\sim 2\times10^5$ and leads to the energy shift displayed in panel (b).
\label{SO_Displacement}}
\end{figure*}

Using eqn.(1) of the main text, as shown in Fig.4, the emitted spectra have been fitted yielding a dimensionless parameter $\chi=0.65$. The driving angular frequency $\omega_d$ corresponds to the coherent mechanical cavity confined fundamental breathing mode ($\hbar\omega_d\simeq 80\mu eV$), which induces an energy shift of the optical mode $\Delta E_{BEC}=\hbar\Delta\omega_0=\hbar\omega_d~ \chi \sim 55\mu eV$.

The presence of the confined acoustic mode, induces a shift of the interfaces of the structure, and this expansion/contraction modifies the interaction with the electromagnetic fields, i.e. with the cavity exciton-polatiton system. In order to estimate the average number of acoustic phonons $\langle N\rangle$ driving the system, together with the associated strain, we proceed the following way.

The electromagnetic fields are calculated assuming a planar structure, using a conventional transfer matrix formalism (TMF), including a complex dielectric susceptibility that accounts for the excitonic states of the embedded quantum wells. Based on the TMF, the optical reflectivity spectrum is calculated and depicted in Fig.~S\ref{SO_Displacement}(a). Here, both resulting modes (lower and upper polatiton modes, LP and UP) are observed, and represent a situation of a slight negative cavity-exciton detuning.

The approach used is quasi-static, i.e. the optical frequencies are high enough compared to the frequencies of the phononic system to consider the vibrations as ``frozen''. The cavity mechanical acoustic vibrations are calculated using an elastic continuum one-dimensional model, also using conventional TMF, and considering the corresponding elastic boundary conditions and the mode's normalization. The calculation for the cavity confined fundamental breathing mode is shown in Fig.~S\ref{SO_Displacement}(c). The acoustic modes displacement (in red) is superimposed to the profile of the structure's index of refraction. The region of the cavity is shaded in grey in between the top/bottom DBRs. Here, $z=0$ corresponds to the air-sample surface. Figure~S\ref{SO_Displacement}(d), shows a detail  of the cavity spacer's region of Fig.~S\ref{SO_Displacement}(c). As observed, the vibrational mode induces a symmetric expansion-contraction of the spacer (vertical blue arrows).

 The value of the acoustic displacement field $U(z)$ at the interfaces between the different materials indicate the modification of the un-perturbed structure. This modification of the structure given by the effective expansion/contraction of the layers is considered for calculating the perturbed optical reflectivity spectrum. In Fig.~S\ref{SO_Displacement}(b) a zoom of the LP cavity mode is presented. As observed, the action of the acoustic strain (in its contraction state) is to shift the optical mode to higher energies. To estimate the polariton mode's energy shift $\Delta E_{BEC}$, we increased the number of phonons $\langle N\rangle$ populating this mechanical confined mode, to match the experimentally estimated energy shift of $55\mu eV$ [see Fig.~S\ref{SO_Displacement}(b)]. The estimated number corresponds to $\langle N\rangle\sim 2\times10^5$, and the resulting confined mode is shown in S\ref{SO_Displacement}(d), where the phonon's amplitude at each cavity interface is $\sim 330$ pm. Given the fact that the nominal cavity width is $d_{cav}=590$ nm, the associated strain is of the order of 0.1\%.

\section{Radiation Pressure (RP) vs. Electrostriction (ES) spectra in non-etched / etched sample regions}

\begin{figure*}[!hht]
    \begin{center}
    \includegraphics[clip=true, keepaspectratio=true, width=1\columnwidth, angle=0]{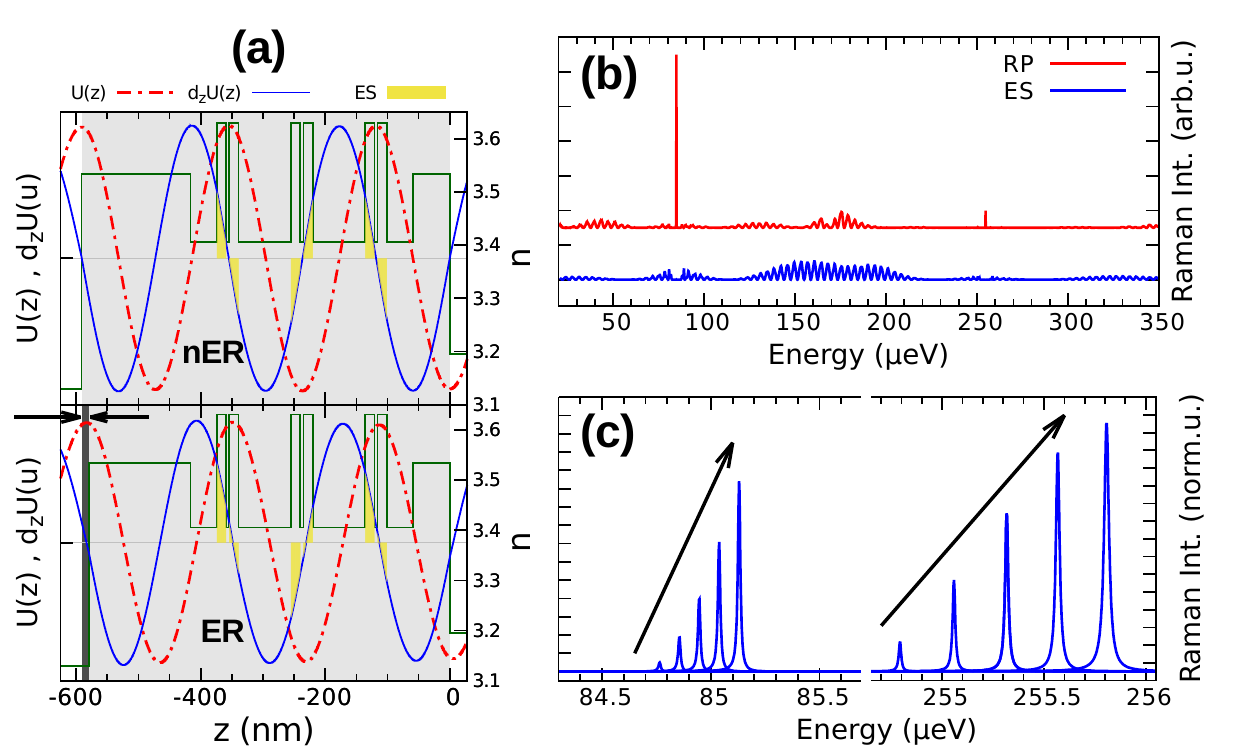}
    \end{center}
    \vspace{-0.5 cm}
\caption{
\textbf{Vibrational spectra due to radiation pressure and electrostriction optical forces.}
Panel (a): Acoustic displacement field $U(z)$ (dashed-dotted red curve) of the cavity confined mechanical modes, together with their associated strain $d_zU(z)$ (blue curves). The two regions of the sample are analyzed, non-etched (nER), and etched regions (ER).
The profile of the index of refraction ($n$, dark-green lines) shows the detail of the cavity spacer (gray-shaded area) of the structure.
Panel (b): Vibrational spectra of both radiation pressure (RP, red curve) and electrostriction (ES, blue curve) forces, corresponding to the non-etched regions.
Panel (c): Evolution of the Raman polarizability due to electrostriction interaction when the cavity spacer is increasingly etched.
\label{RPvsES}}
\end{figure*}

We discuss in this section the amplitude of the different cavity vibrational modes depending on the mechanism of light-matter interaction, namely radiation pressure and exciton-mediated electrostriction. The calculations were performed on a model structure with layer thicknesses somewhat smaller than the precise microstructure reported in the paper. This explains the slightly larger magnitude of the calculated vibrational frequencies, when compared to the experiments reported in the paper. The models used are the same as described in Ref.~\cite{Villafane1}.

In Fig.~S\ref{RPvsES}(a) the profile of the index of refraction ($n$, dark-green lines) shows the detail of the cavity spacer (gray-shaded area) of the structure. The two regions of the sample are analyzed: The top panel corresponds to the non-etched region (nER), while the bottom panel to the etched region (ER), i.e. to the region surrounding the cavity polariton traps. Superimposed for both cases, the acoustic displacement field $U(z)$ (dashed-dotted red curve) of the cavity confined mechanical modes are shown, together with their associated strain $d_zU(z)$ (blue curves). The region that is edged out (bottom panel) is indicated by the horizontal black arrows. The $U(z)$ of the nER (top panel), is the same field displayed in Fig.~S\ref{SO_Displacement}. One important point to notice here is the fact that for the nER the mechanical modes' antinodes fall precisely in between the quantum well (QW) pairs. As a consequence, the associated strain $d_U(z)$ has a node there. As we will explain next, this fact has drastic consequences for one of the polariton-phonon interaction mechanisms, i.e. for the electrostriction (ES) process. This interaction, is basically proportional to the integrated strain present at the QWs (region marked with yellow). As can be observed, due to the fields' symmetry with respect to the QWs, for the nER, the areas contributing positively are canceled out by those contributing negatively. While the radiation pressure process (RP) is only influenced by the effective expansion/contraction of the cavity, which is only very weakly affected by the etching, for electrostriction (ES) at the polariton traps (nER) the effect is practically zero. Interestingly, when the etching takes on, the symmetry brakes down and the strain for the QWs is modified yielding a non-zero effective  contribution.

In Fig.~S\ref{RPvsES}(b) we compare the contributions to the vibrational spectra of both radiation pressure (RP, red curve) and electrostriction (ES, blue curve) corresponding to the non-etched regions.  These curves are obtained from the corresponding overlap integral between the initial and scattered electromagnetic field \cite{Villafane1}, which is proportional to the Raman polarizability. As can be seen, RP shows a very strong coupling to the confined modes at $\sim$22\,GHz ($\sim 85\,\mu$eV) and at $\sim 66\,$GHz($\sim 255\,\mu$eV), while the contribution for ES is absent (no peaks can be observed at these frequencies).

To show the increasing coupling when passing from one region to the other, in Fig.~S\ref{RPvsES}(c) we analyze the evolution of the Raman polarizability due to electrostriction interaction when the cavity spacer is progresively etched. It can be well seen that the coupling via electrostriction increases rapidly and significantly ($\sim\times 30$) when the spacer is reduced (following the black arrows), affecting the coupling to the fundamental $\nu_m^0$ and the second overtone $\nu_m^1$. The etching steps in Fig.~S\ref{RPvsES}(c) are 2\,nm, starting from the nominal non-etched structure (nER in Fig.~S\ref{RPvsES}(a)) and towards the situation described in Fig.~S\ref{RPvsES}(a) bottom panel (ER). Notice that in fact the vibrational mode at $\sim$66\,GHz($\sim 240\,\mu$eV) actually couples more efficiently to the cavity polariton via ES than the fundamental mechanical mode.

\section{Two-polariton mode Hamiltonian coupled to linear and quadratic optomechanical interactions}

The Hamiltonian
\be
\Hop_0=\Hop_{\traptop}+\Hop_{\traplow}+\hbar\phononFreq{} \phononOpD \phononOp +\Hop_\mathrm{int}
\ee
describes as isolated the two traps coupled to a phonon mode of frequency $\phononFreq{}$ with bosonic annihilation (creation) operator $\phononOp$ ($\phononOpD$). The traps Hamiltonians are
\be
{\Hop}_i=\hbar\omega_{e,i}\excitonOp^\dagger_i\excitonOp_i+\hbar\omega_{c,i}\photonOp^\dagger_i\photonOp_i+\frac{1}{2}\hbar\xi_i(\photonOp^\dagger_i\excitonOp_i+\excitonOp^\dagger_i\photonOp_i)
\ee
where $\excitonOp_i$ ($\excitonOp_i^\dagger$) and $\photonOp_i$ ($\photonOp_i^\dagger$) are the bosonic annihilation (creation) operators of the cavity exciton and cavity photon fields, respectively.  In each trap $\xi_i$ accounts for the exciton-photon coupling and $\omega_{e,i}$ ($\omega_{c,i}$) accounts for the cavity exciton (photon) frequency.

The interaction between the traps and the phonon mode includes a radiation pressure and an electrostriction contribution through the coupling strengths $g_{0,rp}$ and $g_{0,es}$, respectively.  We write the interaction as
 \be
\Hop_\mathrm{int}= -\hbar g_{0,rp}  \photonOp_\traplow^\dagger \photonOp_\traptop \phononOpD -\hbar g_{0,es} \excitonOp_\traplow^\dagger \excitonOp_\traptop \phononOpD +h.c.
\ee
where we assumed that the trap \traptop's polariton BEC mode is blue-shifted with respect to trap \traplow's one and followed our measurements insight that the Stokes process is due to mixing of two neighbor trap modes in the optomechanic coupling.

Using the polaritons' basis each trap Hamiltonian becomes diagonal,
\be
 {\Hop}_i=\hbar\Omega_{i,+}\polaritonOp_{i,+}^{\dagger}\polaritonOp_{i,+}+\hbar\Omega_{i,-}\polaritonOp_{i,-}^{\dagger}\polaritonOp_{i,-}
\ee
 where $2\Omega_{i,\pm}=\omega_{c,i}+\omega_{e,i}\pm\sqrt{(\omega_{c,i}-\omega_{e,i})^2+\xi_i^2}$ and the polariton annihilation operators are, $\polaritonOp_{i,+}=\cos\theta_i \photonOp_i+\sin\theta_i \excitonOp_i$ and $\polaritonOp_{i,-}=\cos\theta_i \excitonOp_i-\sin\theta_i \photonOp_i$ with $2\theta_i=\arctan(\xi_i/(\omega_{c,i}-\omega_{e,i}))$.  For $\omega_{c,i}=\omega_{e,i}$ one has $\theta_i=\pi/4$ i.e., the polaritons' excitonic and photonic components are equal. On the other hand for $\omega_{c,i}\ll\omega_{e,i}$ one has $\theta_i\approx\pi/2$ and the low polariton (LP) solution becomes photonic: $\polaritonOp_{i,-}\approx- \photonOp_i$ and $\Omega_{i,-}\approx\omega_{c,i}$.

As traps' BEC modes are due to the LP branches we rewrite $\Hop_\mathrm{int}$ neglecting other polariton branches as:
\be
\Hop_\mathrm{int}= -\hbar G \polaritonOp_{\traplow,-}^\dagger \polaritonOp_{\traptop,-} \phononOpD +h.c.
\ee
with $G=g_{0,rp}  \sin \theta_\traptop  \sin \theta_\traplow + g_{0,es}  \cos \theta_\traptop  \cos \theta_\traplow $.

In what follows we work solely with the LP phonon branches omitting the subindex -, and retaining the trap index $\{\traptop,\traplow \}$. We incorporate a strong effective driving at $\omega_a$ tuned to the blue-shifted polariton mode, of frequency $\Omega_{\traptop}$, and a much weaker effective driving at $\omega_b$  tuned to feeding the lower energy polariton mode of the neighbor trap, of frequency $\Omega_{\traplow}$. We write the optomechanical interaction assuming a situation with $\Omega_\traptop-\Omega_\traplow \gtrsim \phononFreq{}$, making neighbor trap mode mixing in the OM coupling---if nonzero as we assume here---relevant.  The full Hamiltonian reads
\begin{eqnarray*}
\Hop &=&\hbar\phononFreq{} \phononOpD  \phononOp  + \hbar\Omega_{\traptop}\polaritonOp_{\traptop}^{\dagger}\polaritonOp_{\traptop} + \hbar\Omega_{\traplow}\polaritonOp_{\traplow}^{\dagger}\polaritonOp_{\traplow}   -(\hbar G \polaritonOp_{\traplow}^\dagger \polaritonOp_{\traptop} \phononOpD +h.c)\nonumber\\
&&+\ci\hbar\eta_{a,\traptop}\left(  \mathrm{e}^{-\ci\omega_a t} \polaritonOp_{\traptop}^\dagger-\mathrm{e}^{\ci\omega_a t} \polaritonOp_{\traptop}\right) +\ci\hbar\eta_{b,\traplow}\left(  \mathrm{e}^{-\ci\omega_b t} \polaritonOp_{\traplow}^\dagger-\mathrm{e}^{\ci\omega_b t} \polaritonOp_{\traplow}\right)   + \Hop_\mathrm{decay}.
\end{eqnarray*}
 Cavity losses are included in $\Hop_\mathrm{decay}$ parametrized by the polariton traps' linewidths $\kappa_\traptop$, $\kappa_\traplow$ and the phonon linewidth $\Gamma_m$.  The driving amplitudes are controlled by the rates $\eta_{a,\traptop}$ and $\eta_{b,\traplow}$.\cite{Note}

The customary approach of working in the interaction picture and linearizing the equations of motion around equilibrium  (for classical complex amplitudes) leads to--after solving for the perturbations in Fourier space-- the optomechanically modified phonon effective lifetime $\Gamma_{eff}=\Gamma_m (1-C)$. This means that the threshold for self-oscillation is reached provided
\be
1<C=4 \frac{N_{\traptop} |G|^2}{\kappa_{2}\Gamma_m}.
\ee
The latter result holds for  $\Omega_\traplow=\omega_b$, and $\Omega_\traptop-\Omega_\traplow=\omega_m$;  for the strongest driven trap mode the optomechanical coupling has been neglected: Under the undepleted pump approximation, at $\Omega_\traplow=\omega_a$, this mode is assumed having a time-independent large population of $N_\traptop\propto |\eta_{a,\traptop}|^2$.

The experimental results indicate that higher-order terms in the phonon operators could also be present in $\Hop_\mathrm{int}$. For example a quadratic interaction of the form $(-\hbar G_2 \polaritonOp_{\traplow}^\dagger \polaritonOp_{\traptop} (\phononOpD)^2 +h.c)$ could be the most important contribution when the traps are detuned by  $\Omega_\traptop-\Omega_\traplow=2\omega_m$. In general,  for these higher-order couplings the conditions for achieving the threshold---even when non-detuned---depend nontrivially of the state of the system as the number of cavity phonons and the numbers of polaritons in each trap. This can be seen already for the pure quadratic case with $G=0$ and $G_2\neq 0$.
The exploration of any particular higher-order case is beyond the scope of this work.\\


\begin{thebibliography}{36}

\bibitem{Kurizki} G. Kurizki, P. Bertet, Y. Kubo, K. M{\o}lmer, D. Petrosyan, D., P. Rabl, J. Schmiedmayer, Quantum technologies with hybrid systems, {\it Proceedings of the National Academy of Sciences} {\bf 112}, 3866 (2015).

\bibitem{ReviewCOM} M. Aspelmeyer, T. J. Kippenberg, and F. Marquardt, Cavity optomechanics, {\it Rev. Mod. Phys.} {\bf 86}, 1391 (2014).

\bibitem{Rogers} B. Rogers, N. L.  Gullo, G. De Chiara, G. M. Palma, M. Paternostro, Hybrid optomechanics for quantum technologies, {\it Quantum Measurements and Quantum Metrology} {\bf 2}, 11 (2014).


\bibitem{Kippenberg} T. J. Kippenberg, H. Rokhsari, T. Carmon, A. Scherer, K. J. Vahala, Analysis of Radiation-Pressure Induced Mechanical Oscillation of an Optical Microcavity, {\it Phys. Rev. Lett.} {\bf 95}, 033901 (2005).

\bibitem{Grudinin} I. S. Grudinin, H. Lee, O. Painter, and K. J. Vahala, Phonon Laser Action in a Tunable Two-Level System,  {\it Phys. Rev. Lett.} {\bf 104}, 083901 (2010).



\bibitem{O'Connell} A. D. O'Connell, M. Hofheinz, M. Ansmann, Radoslaw C. Bialczak, M. Lenander,
Erik Lucero, M. Neeley, D. Sank, H. Wang, M. Weides, J. Wenner, John M. Martinis, and A. N. Cleland, Quantum ground state and single-phonon control of a mechanical resonator, {\it Nature} {\bf 464}, 697 (2010).

\bibitem{Teufel} J. D. Teufel, T. Donner, Dale Li, J. H. Harlow, M. S. Allman, K. Cicak, A. J. Sirois, J. D. Whittaker, K. W. Lehnert, and R. W. Simmonds, Sideband cooling of micromechanical motion to the quantum ground state, {\it Nature} {\bf 475}, 359 (2011).

\bibitem{Chan} J. Chan, T. P. Mayer Alegre, Amir H. Safavi-Naeini, Jeff T. Hill, Alex Krause, Simon Groeblacher, Markus Aspelmeyer, and Oskar Painter, Laser cooling of a nanomechanical oscillator into its quantum ground state, {\it Nature} {\bf 478}, 89 (2011).

\bibitem{Verhagen} E. Verhagen, S. Deleglise, S. Weis, A. Schliesser, and T. J. Kippenberg, Quantum-coherent coupling of a mechanical oscillator to an optical cavity mode, {\it Nature} {\bf 482}, 63 (2012).


\bibitem{Bochmann} J. Bochmann, A. Vainsencher, D. D. Awschalom, A. N. Cleland, Nanomechanical coupling between microwave and optical photons, {\it Nature Physics} {\bf 9}, 712 (2013).

\bibitem{Bagci} T. Bagci, A. Simonsen, S. Schmid, L. G. Villanueva, E. Zeuthen, J. Appel, J. M. Taylor, A. Sorensen, K. Usami, A. E. Schliesser, S. Polzik, Optical detection of radio waves through a nanomechanical transducer, {\it Nature} {\bf 507}, 81 (2014).

\bibitem{Andrews} R. W. Andrews, R. W. Peterson, T. P., Purdy, K. Cicak, R. W. Simmonds, C. A. Regal, K. W. Lehnert, Bidirectional and efficient conversion between microwave and optical light, {\it Nature Physics} {\bf 10}, 321 (2014).


\bibitem{Weisbuch} C. Weisbuch, CM. Nishioka, A. Ishikawa, Y. Arakawa, Observation of the coupled exciton-photon mode splitting in a semiconductor quantum microcavity. {\it Physical Review Letters} {\bf 69}, 3314 (1992).

\bibitem{Bajoni} D. Bajoni, P. Senellart, E. Wertz, I. Sagnes, A. Miard, A. Lema{\'i}tre, J. Bloch,  Polariton laser using single micropillar GaAs- GaAlAs semiconductor cavities, {\it Physical Review Letters} {\bf 100}, 047401 (2008).

\bibitem{Kasprzak} J. Kasprzak, M. Richard, S. Kundermann, A. Baas, P. Jeambrun, J. M. J. Keeling, F. M. Marchetti, M. H. Szymanska, R. Andr\'e, J. L. Staehli,  V. Savona,  Bose-Einstein condensation of exciton polaritons, {\it Nature} {\bf 443}, 409 (2006).

\bibitem{Amo} A. Amo, J. Lefr{\'e}re, S. Pigeon, C. Adrados, C. Ciuti, I. Carusotto, R Houdr\'e, E. Giacobino,  A. Bramati,  Superfluidity of polaritons in semiconductor microcavities, {\it Nature Physics} {\bf 5}, 805 (2009).

\bibitem{Schneider} C. Schneider, A. Rahimi-Iman, N. Y. Kim, J. Fischer, I. G. Savenko, M. Amthor, M. Lermer, A. Wolf, L. Worschech, V. D. Kulakovskii,  I. A.  Shelykh, An electrically pumped polariton laser, {\it Nature} {\bf 497}, 348 (2013).

\bibitem{Rodriguez} S. R. K. Rodriguez, A. Amo, I. Sagnes, L. Le Gratiet, E. Galopin, A. Lemaitre, J. Bloch, Interaction-induced hopping phase in driven-dissipative coupled photonic microcavities, {\it Nature Communications} {\bf 7}, 11887 (2016).


\bibitem{Restrepo} J. Restrepo, C. Ciuti, I. Favero,  Single-polariton optomechanics, {\it Physical Review Letters} {\bf 112}, 013601 (2014).

\bibitem{Kyriienko} O. Kyriienko, T. C. H. Liew, I. A. Shelykh,  Optomechanics with cavity polaritons: dissipative coupling and unconventional bistability, {\it Physical Review Letters} {\bf 112}, 076402 (2014).


\bibitem{Jusserand} B. Jusserand, A. N. Poddubny, A. V. Poshakinskiy, A. Fainstein, A. Lemaitre,  Polariton resonances for ultrastrong coupling cavity optomechanics in GaAs/AlAs multiple quantum wells. {\it Physical Review Letters}, {\bf 115}, 267402 (2015).


\bibitem{Brennecke} F. Brennecke, S. Ritter, T. Donner, T. Esslinger, Cavity optomechanics with a Bose-Einstein condensate, {\it Science} {\bf 322}, 235 (2008).

\bibitem{FainsteinPRL2013} A. Fainstein, N. D. Lanzillotti-Kimura, B. Jusserand, B. Perrin, Strong optical-mechanical coupling in a vertical GaAs/AlAs microcavity for subterahertz phonons and near-infrared light, {\it Physical Review Letters} {\bf 110}, 037403 (2013).

\bibitem{Anguiano} S. Anguiano, A. E. Bruchhausen, B. Jusserand, I. Favero, F. R. Lamberti, L. Lanco, I. Sagnes, A. Lemaitre, N. D. Lanzillotti-Kimura, P. Senellart, A. Fainstein, Micropillar resonators for optomechanics in the extremely high 19-95-GHz frequency range, {\it Physical Review Letters} {\bf 118}, 263901 (2017).



\bibitem{Kuznetsov} A. S. Kuznetsov, P. L. Helgers, K. Biermann, P. V.  Santos,  Quantum confinement of exciton-polaritons in a structured (Al, Ga) As microcavity,  {\it Physical Review B} {\bf 97}, 195309 (2018).



\bibitem{Renninger} P. Kharel, G. I. Harris, E. A. Kittlaus, W. H. Renninger, N. T. Otterstrom, J. G. E. Harris, P. T. Rakich, High-frequency cavity optomechanics using bulk acoustic phonons, {\it Science Advances} {\bf 5}, eaav0582 (2019).


\bibitem{FainsteinPRL1995} A. Fainstein, B. Jusserand, V. Thierry-Mieg, Raman scattering enhancement by optical confinement in a semiconductor planar microcavity, {\it Physical Review Letters} {\bf 75}, 3764 (1995).




\bibitem{RozasRSI}  G. Rozas, B. Jusserand, A. Fainstein, Fabry-P{\'e}rot-multichannel spectrometer tandem for ultra-high resolution Raman spectroscopy, {\it Review of Scientific Instruments} {\bf 85}, 013103 (2014).


\bibitem{QDs} M. Metcalfe, S. M. Carr, A. Muller, G. S. Solomon, J. Lawall,  Resolved sideband emission of InAs/GaAs quantum dots strained by surface acoustic waves, {\it Physical Review Letters} {\bf 105}, 037401 (2010).

\bibitem{NVs} D. A. Golter, T. Oo, M. Amezcua, K. A. Stewart, H. Wang, Optomechanical quantum control of a nitrogen-vacancy center in diamond, {\it Physical Review Letters} {\bf 116}, 143602 (2016).


\bibitem{Note} Notice that the transition to the self-oscillation regime could be controlled  either by tuning the neighbors' adequate energy level by the proper design of the system, so that their detuning is resonant with the mechanical frequency at a certain power, or by including an additional non-resonant pump laser to independently fine-tune the neighbors fundamental modes energy.


\bibitem{Perrin} C. Lagoin, B. Perrin, P. Atkinson, D. Garcia-Sanchez, High spectral resolution of GaAs/AlAs phononic cavities by subharmonic resonant pump-probe excitation, {\it Physical Review B} {\bf 99},  060101 (2019).


\bibitem{Villafane1} V. Villafa\~ne, P. Sesin, P. Soubelet, S. Anguiano, A. E. Bruchhausen, G. Rozas, C. Gomez Carbonell, A. Lema{\'i}tre, and A. Fainstein, Optoelectronic forces with quantum wells for cavity optomechanics in GaAs/AlAs semiconductor microcavities, {\it Physical Review B} {\bf 97}, 195306 (2018).



\bibitem{Huyhn} A. Huynh, N. D. Lanzillotti-Kimura, B. Jusserand, B. Perrin, A. Fainstein, M. F. Pascual-Winter, E. Peronne, and A. Lema{\'i}tre, Subterahertz phonon dynamics in acoustic nanocavities, {\it Physical Review Letters} {\bf 97}  115502 (2006).


\bibitem{Rueda} A. Rueda, F. Sedlmeir, M. C. Collodo, U. Vogl, B. Stiller, G. Schunk, D. V. Strekalov, C. Marquardt, J. M. Fink, O. Painter, G. Leuchs, Efficient microwave to optical photon conversion: an electro-optical realization, {\it Optica} {\bf 3}, 597 (2016).


\bibitem{Villafane2} V. Villafa\~ne, S. Anguiano, A. E. Bruchhausen, G. Rozas, J. Bloch, C. Gomez Carbonell, A. Lema{\'i}tre, and A. Fainstein, Quantum well photoelastic comb for ultra-high frequency cavity optomechanics, {\it Quantum Science and Technology} {\bf 4}, 014011 (2018).





\end{thebibliography}

\begin{thebibliography}{7}


\bibitem{Kuznetsov} Kuznetsov, A. S., Helgers, P. L., Biermann, K., and  Santos, P. V., Quantum confinement of exciton-polaritons in a structured (Al, Ga) As microcavity,  Physical Review B {\bf 97}, 195309 (2018).




\bibitem{RozasRSI}  Rozas, G., Jusserand, B., and Fainstein, A., Fabry-P\'erot-multichannel spectrometer tandem for ultra-high resolution Raman spectroscopy, Review of Scientific Instruments {\bf 85}, 013103 (2014).


\bibitem{Johnson2002}
S.~G. Johnson, M.~Ibanescu, M.~A. Skorobogatiy, O.~Weisberg, J.~D. Joannopoulos, and Y.~Fink, Perturbation theory for Maxwell's equations with shifting material boundaries, Phys. Rev. E \textbf{65}, 066611 (2002).

\bibitem{Ding2010} L.~Ding, C.~Baker, P.~Senellart, A.~Lema\^itre, S.~Ducci, G.~Leo, and I.~Favero, High frequency {G}a{A}s nano-optomechanical disk resonator, Phys. Rev. Lett. \textbf{105}, 263903 (2010).

\bibitem{Baker} C. Baker, W. Hease, Dac-Trung Nguyen, A. Andronico, S. Ducci, G. Leo, and I. Favero, Photoelastic coupling in gallium arsenide optomechanical disk resonators, Optics Express {\bf 22}, 14072 (2014).


\bibitem{Villafane1} Villafa\~ne, V., P. Sesin, P. Soubelet, S. Anguiano, A. E. Bruchhausen, G. Rozas, C. Gomez Carbonell, A. Lema\^itre, and A. Fainstein, Optoelectronic forces with quantum wells for cavity optomechanics in GaAs/AlAs semiconductor microcavities, Physical Review B {\bf 97}, 195306 (2018).


\bibitem{Note}For a source of frequency $\omega_{\alpha}$, power $P_\alpha$ and with the cavity $i$ being able to accept excitations from the source at the rate $\kappa_i^\mathrm{ext}$, one gets  $|\eta_{\alpha,i}|^2=\frac{P_{\alpha}\kappa_i^\mathrm{ext}}{\hbar\omega_\alpha}$.









\end{thebibliography}


\end{document}